\documentclass[aps,showpacs,showkeys]{revtex4}
\usepackage{amsmath,amssymb,amsfonts,epsfig,graphicx,color}

\begin{document}

\title{Community Structure in the United Nations General Assembly}

\author{Kevin T. Macon$^{1,2}$}
\author{Peter J. Mucha$^{1,3}$}\email{mucha@unc.edu}
\author{Mason A. Porter$^{4,5}$}
\affiliation{$^1$Carolina Center for Interdisciplinary Applied Mathematics, Department of Mathematics, University of North Carolina, Chapel Hill, NC 27599-3250, USA}
\affiliation{$^2$Department of Physics, Louisiana State University, Baton Rouge, LA 70803, USA}
\affiliation{$^3$Institute for Advanced Materials, Nanoscience \& Technology, University of North Carolina, Chapel Hill, NC 27599, USA}
\affiliation{$^4$Oxford Centre for Industrial and Applied Mathematics, Mathematical Institute, University of Oxford, OX1 3LB, UK}
\affiliation{$^5$CABDyN Complexity Centre, University of Oxford, Oxford OX1 1HP, UK}

\date{\today}

\begin{abstract}

We study the community structure of networks representing voting on resolutions in the United Nations General Assembly.  We construct networks from the voting records of the separate annual sessions between 1946 and 2008 in three different ways: (1) by considering voting similarities as weighted unipartite networks; (2) by considering voting similarities as weighted, signed unipartite networks; and (3) by examining signed bipartite networks in which countries are connected to resolutions.  For each formulation, we detect communities by optimizing network modularity using an appropriate null model.  We compare and contrast the results that we obtain for these three different network representations.  In so doing, we illustrate the need to consider multiple resolution parameters and explore the effectiveness of each network representation for identifying voting groups amidst the large amount of agreement typical in General Assembly votes.

\end{abstract}



\maketitle


\section{Introduction} \label{intro}

The study of networks has a long history in both the mathematical and social sciences \cite{faust}, and recent investigations have underscored their vibrant interdisciplinary application and development \cite{str01,Newman2003,newman2010,albert,mendes,cald}.  The large-scale organization of real-world networks typically includes coexisting modular (horizontal) and hierarchical (vertical) organizational structures.  Various attempts to interpret such organization have included the computation of structural modules called \textit{communities} \cite{santobig,ourreview,satu07}, which are defined in terms of mesoscopic groups of nodes with more internal connections between nodes in the group than external connections to nodes in other groups. Such communities are not merely structural modules but can have functional importance in network processes.  For example, communities in social networks (``cohesive groups" \cite{moody03}) might correspond to circles of friends or business associates, communities in the World Wide Web might encompass pages on closely-related topics, and some communities in biological networks have been shown to be related to functional modules \cite{amaral,anna}.

As discussed at length in two recent review articles \cite{ourreview,santobig} and references therein, the classes of techniques available to detect communities are both voluminous and diverse.  They include hierarchical clustering methods such as single linkage clustering, centrality-based methods, local methods, optimization of quality functions such as modularity and similar quantities, spectral partitioning, likelihood-based methods, and more.  Investigations of network community structure have been remarkably successful on benchmark examples and have led to interesting insights in several applications, including the role of college football conferences \cite{structpnas} in affecting algorithmic rankings \cite{bcs}; committee assignments \cite{congshort,conglong,conggallery}, legislation cosponsorship \cite{yan}, and voting blocs \cite{waugh} in the U.S.~Congress; the examination of functional groups in metabolic \cite{amaral} and protein interaction \cite{anna} networks; the study of ethnic preferences in school friendship networks \cite{marta}; the organization of online social networks \cite{facebook}, and the study of social structures in mobile-phone conversation networks \cite{jp}.

With the newfound wealth of longitudinal data sets on various types of human activity patterns, it has become possible to investigate the temporal dynamics of communities, and this issue has started to attract an increasing amount of attention \cite{palla07,fenn,fennlong,jp,multislice}.  It is also potentially useful to study community structure in similarity and correlation networks \cite{gomez09}, such as those determined by common voting or legislation-cosponsorship patterns \cite{yan,waugh}, alliances and disputes among nations \cite{traag2009}, or more general coupled time series \cite{fenn,fennlong}.  In such cases, one is faced with numerous choices for how to actually construct the network from the original data---an important issue that has received surprisingly little attention.  (An old discussion of some of the available methods is presented in Ref.~\cite{arend}.)  In the present paper, we focus on this network construction issue by examining the community structure of networks defined in different ways from roll-call voting patterns in the United Nations General Assembly (UNGA).

As discussed in a recent review \cite{burton09}, network analysis has lead to interesting insights in the field of international relations---just as it has in numerous other fields in the social, physical, biological, and information sciences \cite{Newman2003,faust, cald, newman2010}.  For example, elements of the Correlates of War (CoW) data \cite{cow2,cow} have been studied using network methods \cite{cranmer08,traag2009}, and we expect that other available data will similarly admit profitable treatment. Previous studies of UNGA roll-call data have been successful at grouping countries with Nominate scores, which assign ideological coordinates to voting members and can be used to introduce partitions in policy space \cite{voeten,pr97}. Empirical investigation of UNGA voting behavior has become readily accessible due to Voeten's organization of the UNGA voting data \cite{voteworld}. Voeten analyzed this data using Nominate scores \cite{pr97} to study Cold War and post-Cold War voting behavior \cite{voeten}. Lloyd has analyzed similar data, using network and correspondence analysis to show that the so-called ``Clash of Civilizations" does not occur along civilizational lines but rather via a North-South division that arises from economic differences (which she concludes has resulted in different levels of support for human-rights treaties) \cite{paulette}. Motivated by the previously-demonstrated utility of studying community structure in network representations of voting \cite{waugh} and legislation cosponsorship \cite{yan} in the United States Congress, we investigate in this paper the community structure of network representations of the UNGA, based on the patterns of roll-call voting on resolutions, to see if such analysis can help to identify and understand international voting blocs. 

The rest of this paper is organized as follows.  In Section~\ref{data}, we give a brief introduction to the United Nations General Assembly voting data and discuss the different ways that we will represent this data in the form of networks. In particular, we construct (1)
weighted networks defined by the numbers of voting agreements between
pairs of countries; (2) weighted, signed networks in which we separately
consider voting agreements and disagreements between countries; and
(3) signed bipartite networks between countries and resolutions
that directly indicates yes ($+1$) and no ($-1$) votes.  In
Section~\ref{commdetect}, we briefly review community detection via optimization of modularity and its generalizations, emphasizing the use of appropriate
null models with resolution parameters for each of the three kinds of network representations that we consider.  We then investigate community
structure in the UNGA using each of these three formulations and
compare the results that we obtain.  By analyzing the set of
resolutions in each session according to voting agreements
between countries in Section~\ref{agree}, we are able to identify historical trends and changes in the UNGA's community structure.  We then provide case studies by going into further detail in one session from each of three different periods in the UNGA's history.  We investigate these three sessions (the 11th, 36th, and 58th Sessions) in terms of voting agreements and disagreements in
Section~\ref{vote} and as bipartite networks with positive and negative edges in Section~\ref{bip}.  We close with concluding observations in Section~\ref{conc}.


\section{Network Representations of UNGA Voting Data} \label{data}

Unlike the other component bodies of the United Nations (UN), the United Nations General Assembly (UNGA) provides equal representation to all member nations \cite{UNGA}.  Each nation gets one vote, and UNGA representatives can debate international issues and non-binding resolutions.  In recent years, this setting has motivated collaboration among less powerful, developing countries to create a ``North-South'' division.  However, it is unclear how applicable this grouping is in other settings or how cohesive it is on individual issues \cite{russett,paulette,voeten}.  The voting record of UNGA thereby provides an interesting application for the investigation of network community structure. The UNGA roll call also provides a useful setting for testing the effects of different network representations of the data for determining voting blocs, in part because one must pay particularly close attention to the baseline level of agreement that is typical in UNGA resolutions that reach a recorded vote. This baseline tendency towards agreement in the data makes the UNGA roll call different from, e.g., studies of roll-call voting and legislation cosponsorship in the United States Congress \cite{pr97,waugh,yan,fowlershort,fowler}.

\begin{figure*}
  \centerline{\includegraphics[width=0.475\textwidth]{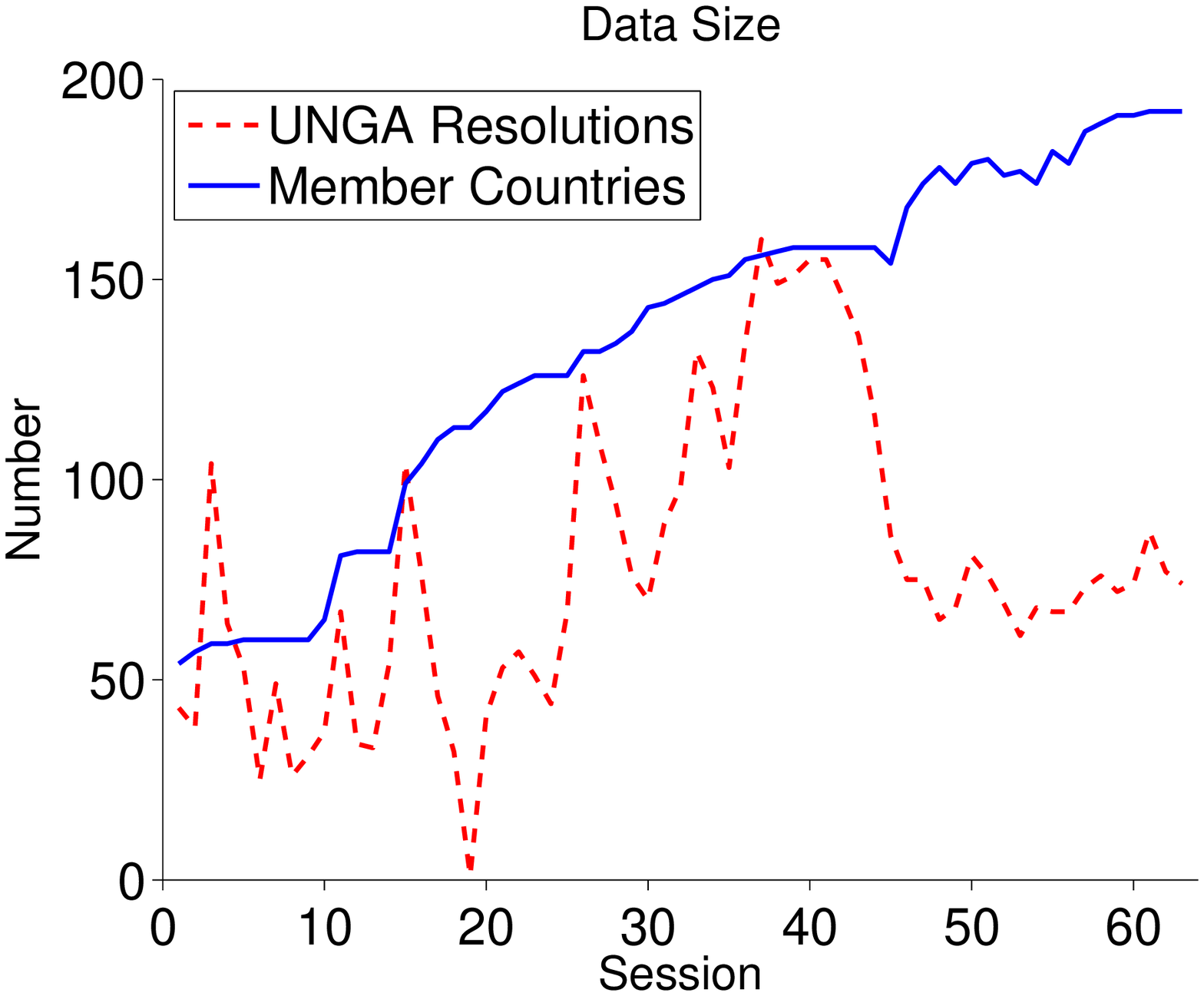}\quad\includegraphics[width=0.475\textwidth]{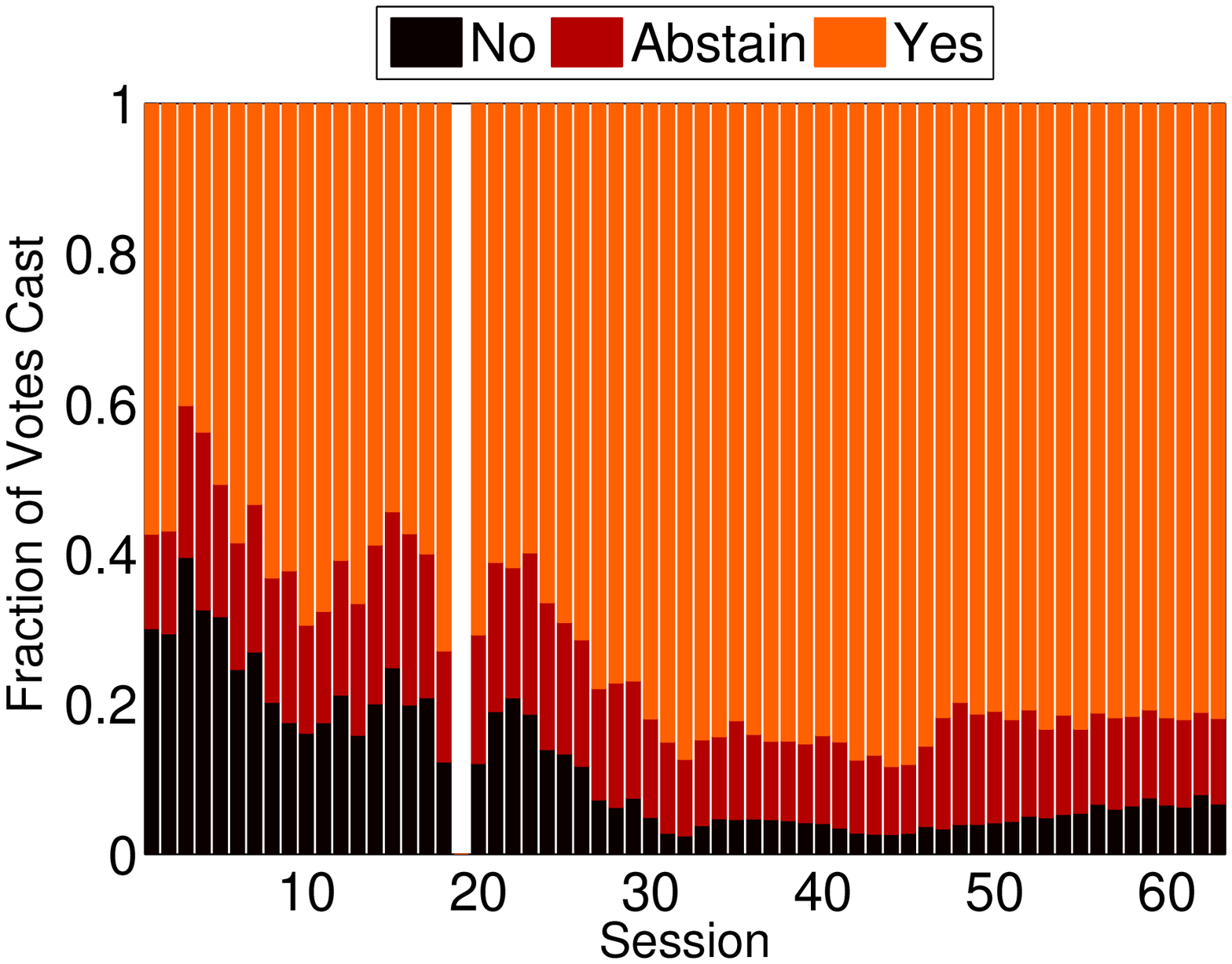}}
  \caption{(Color online) (Left) Numbers of countries (blue) and resolutions (red) in each numbered annual United Nations General Assembly session during 1946--2008 (Session 1 occurred in 1946, etc.). (Right) Fractions of votes cast in each numbered annual session in favor (``yes"), against (``no"), and abstaining.}
  \label{int}
\end{figure*}

The UNGA was founded in 1946.  As indicated in Fig.~\ref{int}, the number of member countries has increased steadily since then, but the number of recorded votes varies from session to session.  In this study, we consider every annual session from 1946 to 2008 except for the 19th session (1964), which we exclude because voting occurred on only one resolution in that session.  We removed unanimous votes from the data, as they do not provide information about the network structure of voting agreements and disagreements between countries. We also note the large amount of agreement in UNGA voting, as illustrated in the right panel of Fig.~\ref{int}, indicating the large fraction of ``yes'' votes among the total set of votes cast. This bias towards agreement skews the simplest measures of voting similarity, so an important area of investigation in the present paper is the consideration of such bias.

Within each session, we define edge weights between pairs of countries using a measure of the level of their voting agreement. For example, Gartzke's ``The Affinity of Nations" data set \cite{gartzke} uses a well-known (in the international relations literature) diagnostic called ``$S$" to measure the relative similarity between UNGA votes ($S$ first converts the voting information to column vectors and then calculates a similarity score between each pair of vectors),
with different calculations depending on whether one includes or omits abstentions \cite{signorino,sweeney}. Assigning numerical values to the types of possible votes---yes, no, abstain, and absent---requires choosing arbitrary relative magnitudes and spacings between these values. For instance, one might assign $\pm 1$ to yes/no votes and choose some intermediate value for abstentions and absences (and, of course, employ some argument as to whether an abstention should be interpreted as closer to a yes or to a no vote, cf.\ similar discussions in Refs.~\cite{arend,voeten}). Contingency table statistics also provide a possible way to measure the agreement between voting countries; they avoid the use of a numerical scale but instead require one to assume an expected distribution of votes \cite{cohen,arend}.

Motivated by Lijhpart's Index of Agreement \cite{arend}, we define a unipartite network of voting similarities in which the strength of connection between a pair of countries is given by the number of agreements on resolutions (yes-yes, no-no, or abstain-abstain). To avoid assigning artificially high agreement scores to countries with low attendance, we do not normalize by the number of times both countries were present and voting.  (This contrasts with the norm when studying voting in legislative networks such as the United States Congress \cite{pr97,congshort,waugh}.)  One might instead uniformly normalize counts of agreement by the total number of votes in a session, so that unit strength indicates perfect agreement on all resolutions in a session.  However, detecting communities on voting agreement networks when using such a constant (in a session) normalization is equivalent to detecting communities on such networks without the normalization. We therefore use the direct count of agreements as the weight of connections in the UN voting agreement networks.  We denote the (weighted, unipartite) adjacency matrix of such a network by $A^+_{ij}$, and we ignore the artificial self-edges imposed by this definition of voting similarity by setting all diagonal entries equal to zero. This is frequently done for correlation networks and is the standard procedure when studying voting similarities \cite{heimo,fenn,fennlong,waugh}.

In contrast with a recent study of Congressional roll call networks \cite{waugh}, the preponderance of ``yes'' votes in the UNGA (see Fig.~\ref{int}) leads to a wealth of large weights in the voting agreement adjacency matrices $A^+_{ij}$, and many of these weights are close to the total number of votes in the session. In this environment of significant agreement in voting results, it is unclear how one should best treat abstentions without additional information that details the causes for each such vote.  In particular, the relative weight of disagreement between two countries in a yes-abstain pair of votes on the same resolution should presumably be treated differently from that in a yes-no pair or in an abstain-no pair.  In order to include the possibility of treating some of these forms of disagreement differently, we count the occasions of direct yes-no disagreement between two countries in the elements of the (symmetric) matrix $A^-_{ij}$ to supplement the $A^+_{ij}$ agreement matrix.  As we discuss in more detail in Section \ref{commdetect}, this will entail performing community detection on signed adjacency matrices.

Finally, to ensure that we are not inappropriately discarding too much information about the particular votes on which countries vote in agreement (or on yes-no disagreement), we will detect communities in a third class of networks.  Each such network (one per UNGA session) consists of a signed bipartite (two-mode) network of countries voting in favor or against individual resolutions. That is, for each session, we will consider the adjacency matrix $\mathbf{V}$ that defines a bipartite network between countries and resolutions that has both positive and negative links. Taking abstentions as zero entries in the absence of other information, we define the matrix elements by
\begin{equation}
	V_{ij}=
\left\lbrace\begin{array}{r l}1,& \text{if country $i$ voted yes on resolution $j$}\\
-1,& \text{if country $i$ voted no on resolution $j$}\\
0,& \text{otherwise (absences \& abstentions).}
\end{array}\right.
\end{equation}
A pair of countries that both vote yes constitutes an actual agreement on a proposed resolution, but there can be multiple reasons for two countries to both vote against a resolution \cite{arend}.  Importantly, the signed bipartite network representation respects this asymmetry.  It is intuitively clear that countries are more likely to have positive edges to resolutions (i.e., yes votes) within their community, but only indirect improvements in group assignment can be obtained by grouping together two countries with negative edges to a given resolution (i.e., no votes).

By investigating community structure in each of these three types of network representations---the $\mathbf{A}^+$ network of agreements alone, the inclusion of both the $\mathbf{A}^+$ agreement network and the $\mathbf{A}^-$ disagreement network, and the underlying bipartite network $\mathbf{V}$ of votes on resolutions---we aim for a more complete and robust picture of the communities present in the UNGA roll call than any one network representation might uncover by itself.  More generally, comparing and contrasting these approaches should provide valuable insight about the network treatment of voting and correlation data.


\section{Community Detection by Optimization of Generalized Modularity} \label{commdetect}

We detect communities by optimizing the quality function known as modularity \cite{structmix,structeval,newmodlong} and some of its generalizations.  (Numerous other graph partitioning methods can of course be employed \cite{ourreview,santobig}.) We take the partition with the highest quality value that we can obtain from among three computational heuristics---spectral bipartitioning \cite{newmod,newmodlong}, spectral tripartitioning \cite{richardson}, and the locally greedy ``Louvain'' method \cite{blondel}---which we subsequently follow in each case by Kernighan-Lin node-swapping steps \cite{kl,newmodlong} in order find a partition of the network that has an even higher value of the quality function.  Of course, other heuristics (including other greedy methods, extremal optimization, and simulated annealing) can also be employed \cite{ourreview,santobig,commreview}.  Because modularity (and any similar quality function) has a complex energy landscape that is expected to include a large number of good local optima, one must take care in interpreting results when using it to study the community structure of real networks \cite{good2010}.  Additionally, modularity optimization has a resolution limit \cite{resolution}, so it is important to include resolution parameters to investigate community structure at multiple scales.  Exploration of the resolution parameter space in each type of network representation is a major focus of the present investigation.

In detecting communities by optimizing modularity (or its generalizations), one partitions a network so that the strength of intra-community edges is optimized relative to a baseline expectation indicated by an appropriate null model.  The quality $Q$ of a partition of the network is a function of the modularity matrix $\mathbf{B}=\mathbf{A}-\gamma\mathbf{P}$, where the adjacency matrix $\mathbf{A}$ encodes the network, the matrix $\mathbf{P}$ encodes the null model, and incorporating a resolution parameter $\gamma$ allows one to identify communities at different scales \cite{santobig,ourreview,spinglass}.  The quality is given by
\begin{equation}
	Q = \sum_{i,j}{B_{ij}\delta(c_i,c_j)}\,,
\end{equation}
where $\delta(c_i,c_j)$ equals $1$ if $i$ and $j$ have been assigned to the same community and $0$ if they have been assigned to different communities.  (The quantity $c_k$ identifies the community to which node $k$ has been assigned.)  Finding the community assignments that maximize $Q$ is an NP-hard problem that requires the use of computational heuristics to obtain a good local optimum~\cite{np,ourreview,santobig}.  Importantly, one must select a null model $\mathbf{P}$ that is appropriate for the network under consideration.  In particular, we need to use a different null model for each of the three different network representations of UNGA voting that we consider (which we recall are the unipartite network of agreement between countries, the unipartite signed network of agreements and disagreements between countries, and the bipartite signed network of yes/no votes by countries on resolutions).

The unipartite network of agreements, encoded by $\mathbf{A}=\mathbf{A}^+$, is the simplest case to consider.  One can employ the standard null model for modularity \cite{structeval}, $P_{ij}=k_ik_j/(2m)$, where $k_i=\sum_j A_{ij}$ is the strength of node $i$ and $2m=\sum_i k_i=\sum_{ij} A_{ij}$ is the sum of all node strengths.  The elements of the modularity matrix are then given by
\begin{equation} \label{Q1}
	B_{ij} = \frac{1}{2m}\left(A_{ij}-\gamma\frac{k_ik_j}{2m}\right)\,.
\end{equation}
When $\gamma = 1$, equation (\ref{Q1}) reduces to the standard definition of modularity \cite{ourreview,structeval}.  The standard null model for modularity describes the expected edge weight in an unsigned, unipartite random graph of independent edges, conditional on having the same expected node strengths as those in the observed network. The standard null model is clearly inappropriate for signed networks and bipartite networks, so other null models must be used instead. In the signed case, the standard null model ignores the potentially important distinction between agreements and disagreements. In the bipartite case, an appropriate null model must respect the constraint that each edge connects nodes of the two different types. 

Graphs with signed edges can be used to study social networks with both sympathetic (positive) and antagonistic (negative) interactions \cite{status}.  This is potentially relevant, for example, for investigations of social balance \cite{renaud10}.  A recent paper introduced a generalization of modularity for signed networks and used it to study a network of international alliances and disputes \cite{traag2009}. In our investigation, we include the $\mathbf{A^{-}}$ representation of yes-no disagreements between countries in addition to the $\mathbf{A^{+}}$ network of agreements.  Recall that the absence of agreements indicated in $\mathbf{A^{+}}$ might include unpaired abstentions or full yes-no disagreement, but we include only the latter of these in the specification of $\mathbf{A^{-}}$ (with such disagreements encoded as $A_{ij}^{-}>0$ elements), providing an opportunity to weight their effects more heavily. Using the signed null model developed in Ref.~\cite{traag2009}, the modularity matrix becomes
\begin{align}
	\nonumber B_{ij} = \frac{1}{2m^{+}+2m^{-}}\left(A^{+}_{ij}-A^{-}_{ij}-\gamma\frac{k^{+}_ik^{+}_j}{2m^{+}}+\lambda\frac{k^{-}_ik^{-}_j}{2m^{-}}\right)\,, \label{uniSigned}
\end{align}
where $k^{\pm}_i=\sum_jA^{\pm}_{ij}$ are the signed strengths for node $i$ and $2m^\pm=\sum_ik_i^{\pm}=\sum_{ij} A_{ij}^{\pm}$ are the corresponding total edge weights of the network's two kinds of edges.  In using two different resolution parameters for agreements and disagreements, one can investigate and separately vary the effects of these two different types of connections on the community structure. In particular, the resolution parameters separately control the expected extent of communities considered along the two kinds of edges (see, e.g., the Laplacian dynamics interpretation of this signed null model in Ref.~\cite{multislice}).

To detect communities in each signed bipartite voting matrix $\mathbf{V}$, we need to generalize the bipartite modularity of Ref.~\cite{barber2007} to incorporate both positive and negative edges. A bipartite network consists of two types of nodes---countries and resolutions---and each edge connects a node of one type to a node of the other.  The matrix $\mathbf{V}$ encodes connections between countries and resolutions.  Ordering the nodes according to type, with all countries ordered before the resolutions, the corresponding modularity matrix $\mathbf{B}$ in bipartite form consists of off-diagonal blocks:
\begin{equation}
	\mathbf{B}=\left[\begin{array}{c c}\mathbf{0}&\mathbf{\tilde{B}}\\ \mathbf{\tilde{B}^{T}} & \mathbf{0} \end{array}\right]\,.
\end{equation}
Recalling our definition of $\mathbf{V}$ with elements $\{0,\pm 1\}$ encoding yes and no votes, the non-zero components of $\mathbf{B}$ then become
\begin{equation} 
	\tilde{B}_{ij}=V_{ij}-\gamma\frac{k^{+}_id^{+}_j}{2m^{+}}+\lambda\frac{k^{-}_id^{-}_j}{2m^{-}}\,,
\end{equation}
where $k_i^{\pm}$ and $d_j^{\pm}$, respectively, denote the positive and negative degrees for country $i$ and resolution $j$, and $2m^{\pm}=\sum_{i}k_i^{\pm}=\sum_{j}d_j^{\pm}$ gives the total number of positive [$(+)$ superscript] and negative [$(-)$ superscript] connections in the network.

Having described the three network representations of the voting record that we will consider, we are faced with the dilemma of setting values for the resolution parameters $\gamma$ and $\lambda$. Given some theoretical justification for expected or desired group sizes, one might reasonably study the partitions obtained for specific resolution values. In the absence of any such additional political information, we will instead explore the space of possible resolutions. In so doing, we will focus on communities that appear robustly across a range of resolution parameter values.


\section{Networks of Vote Agreements} \label{agree}

We first consider the weighted, unipartite UNGA networks that we constructed by considering the level of agreement between counties.  We maximize Newman-Girvan modularity, which is given by equation (\ref{Q1}) with $\gamma=1$, for each of the UNGA sessions.  In addition to its use for community detection, modularity can also be used as a measure of polarization among the voting parties \cite{waugh,yan}.

\begin{figure*}
  \centerline{\includegraphics[width=.475\textwidth]{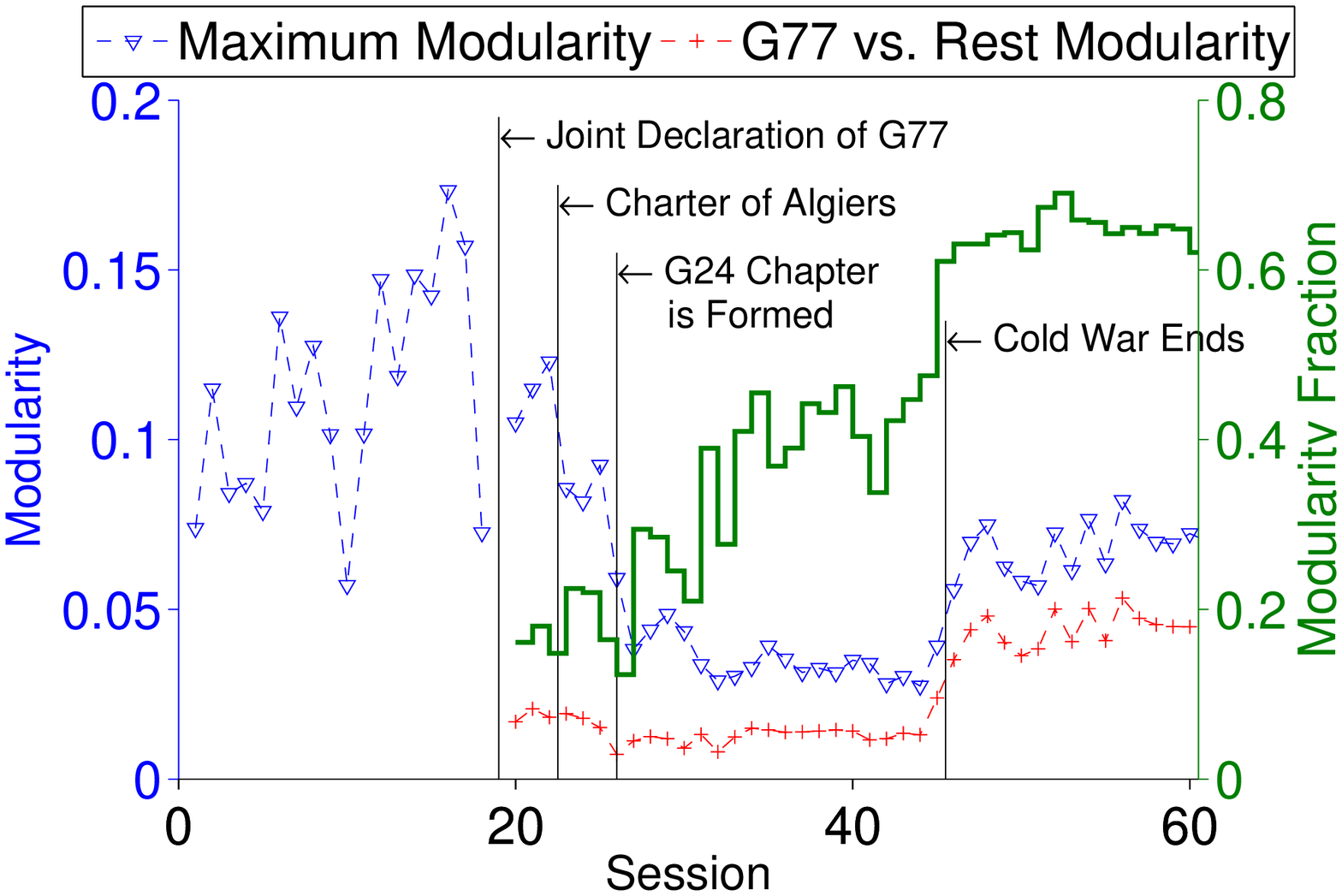}\qquad\qquad\includegraphics[width=.4\textwidth]{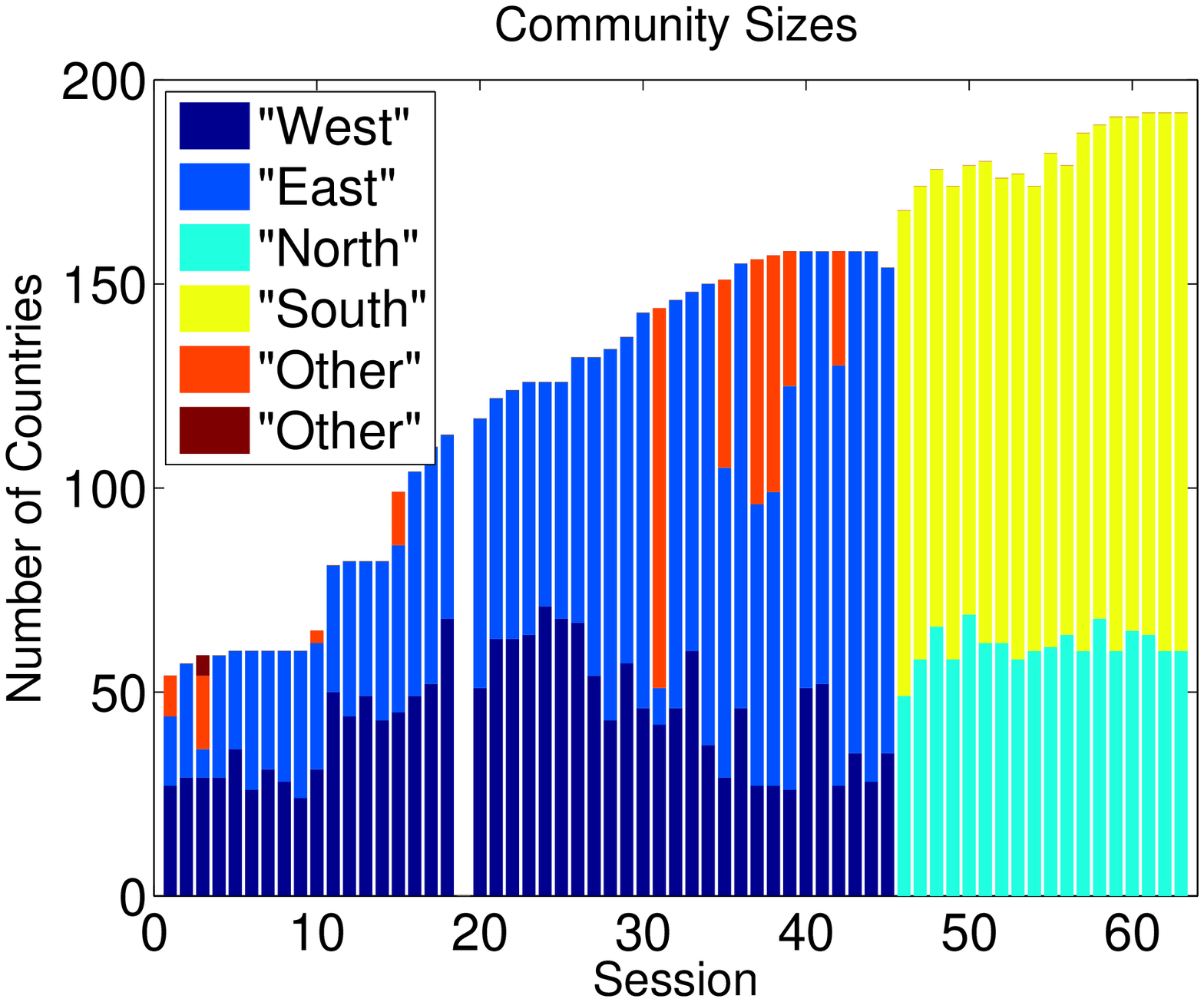}}
  \caption{(Color online) (Left) Maximum modularity obtained by partitioning the (unipartite) voting similarity network of each session. For comparison, we also show the modularity of the partition into G77 and non-G77 countries, revealing the increasing fraction over time of modularity that can be explained by this division of the network.  Observe the sharp increase after the end of the Cold War.
(Right) Communities in each UNGA session that we identified using the network of voting similarities. We labeled the dominant countries geographically in each session by tracking the identities of specific countries (e.g., the United Kingdom is typically West in every session during the Cold War and in the North group after the Cold War).  We list the countries assigned to the North and South communities are in Tables~\ref{NorthBloc} and~\ref{SouthBloc}, respectively.
}
\label{g77}
\end{figure*}

In Fig.~\ref{g77}, we show that beginning near the 1964 declaration of the Group of 77 (G77), the fraction of maximum modularity captured by partitioning on the basis of G77 membership tends to increase over time.  In Fig.~\ref{g77}, we also present the calculated modularity for the two-group G77 partition (G77 members versus non-members \footnote{We identified G77-member countries using current G77 membership status; we did not take into account changes in membership over time.}) for each session. Observe in particular a recent sharp rise in modularity in the Post-Cold War era, with a larger fraction of modularity corresponding to this G77 split.  In the right panel of the figure, we show the sizes of the communities that we found by optimizing the Newman-Girvan modularity in each session, demonstrating that we typically find a small number of large communities. Prior to the end of the Cold War, this dominant split appears to be along an East-West axis, which we identified by tracking the identity of specific countries (e.g., comparing the placement of the UK and the USSR). In contrast, in the Post-Cold War sessions, the two-community split appears to be along a ``North-South'' axis, illustrating the cooperation between developing countries in the dominating North-South division of the UNGA \cite{paulette,russett} (again, e.g., with the UK typically in the North).  In Tables \ref{NorthBloc} and \ref{SouthBloc}, respectively, we list the countries that were consistently assigned to the communities labeled North and South in Sessions 46--63 (1991--2008).  In Table~\ref{SouthBloc}, representing the South, Mexico is the only non-member of the G77 (it left the organization in 1994).  In Table~\ref{NorthBloc}, representing the North, the Marshall Islands is the only member of the G77~\cite{g77}.

\begin{figure*}
  \centerline{\includegraphics[width=.33\textwidth]{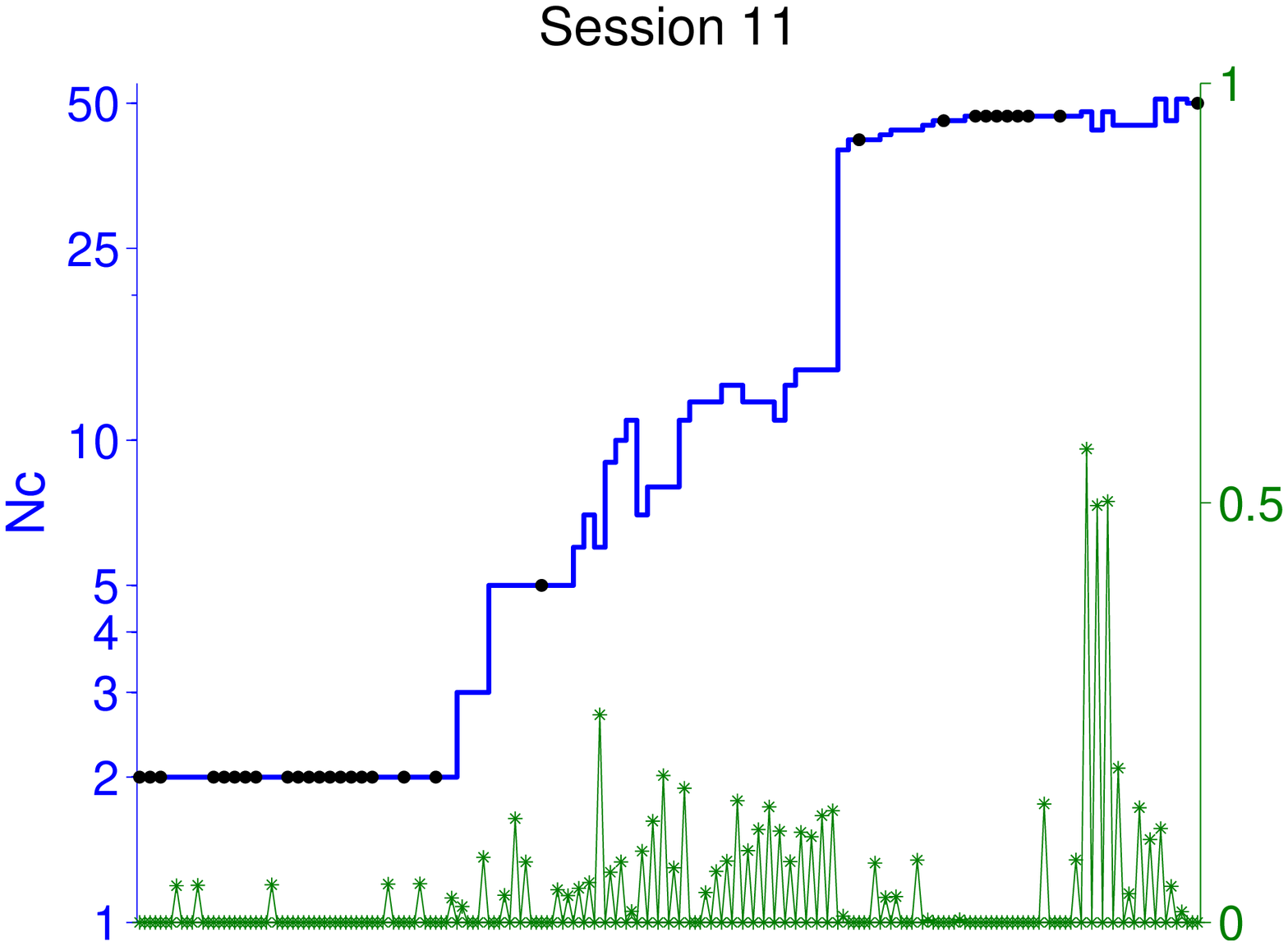}\includegraphics[width=.32\textwidth]{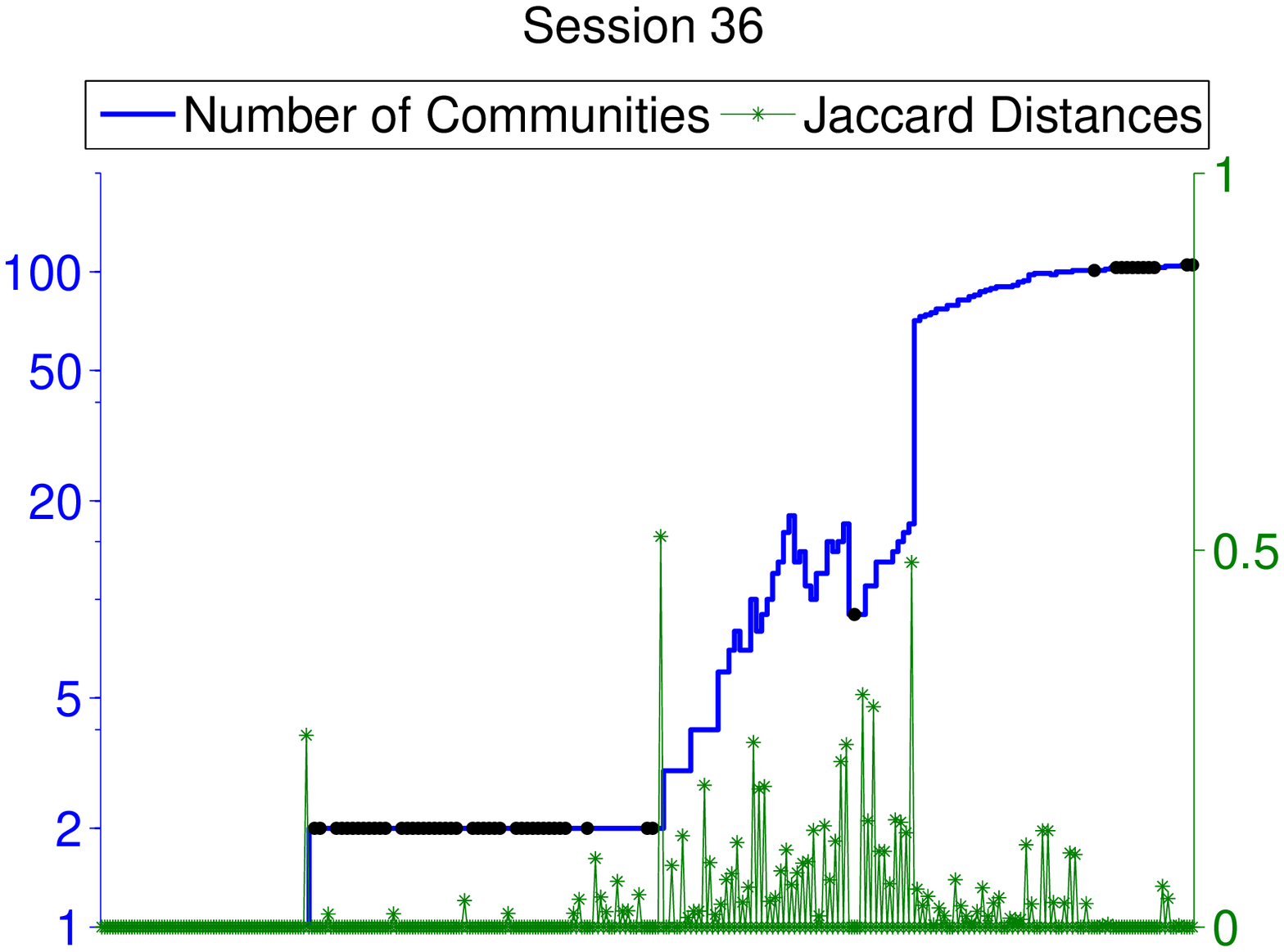}\includegraphics[width=.33\textwidth]{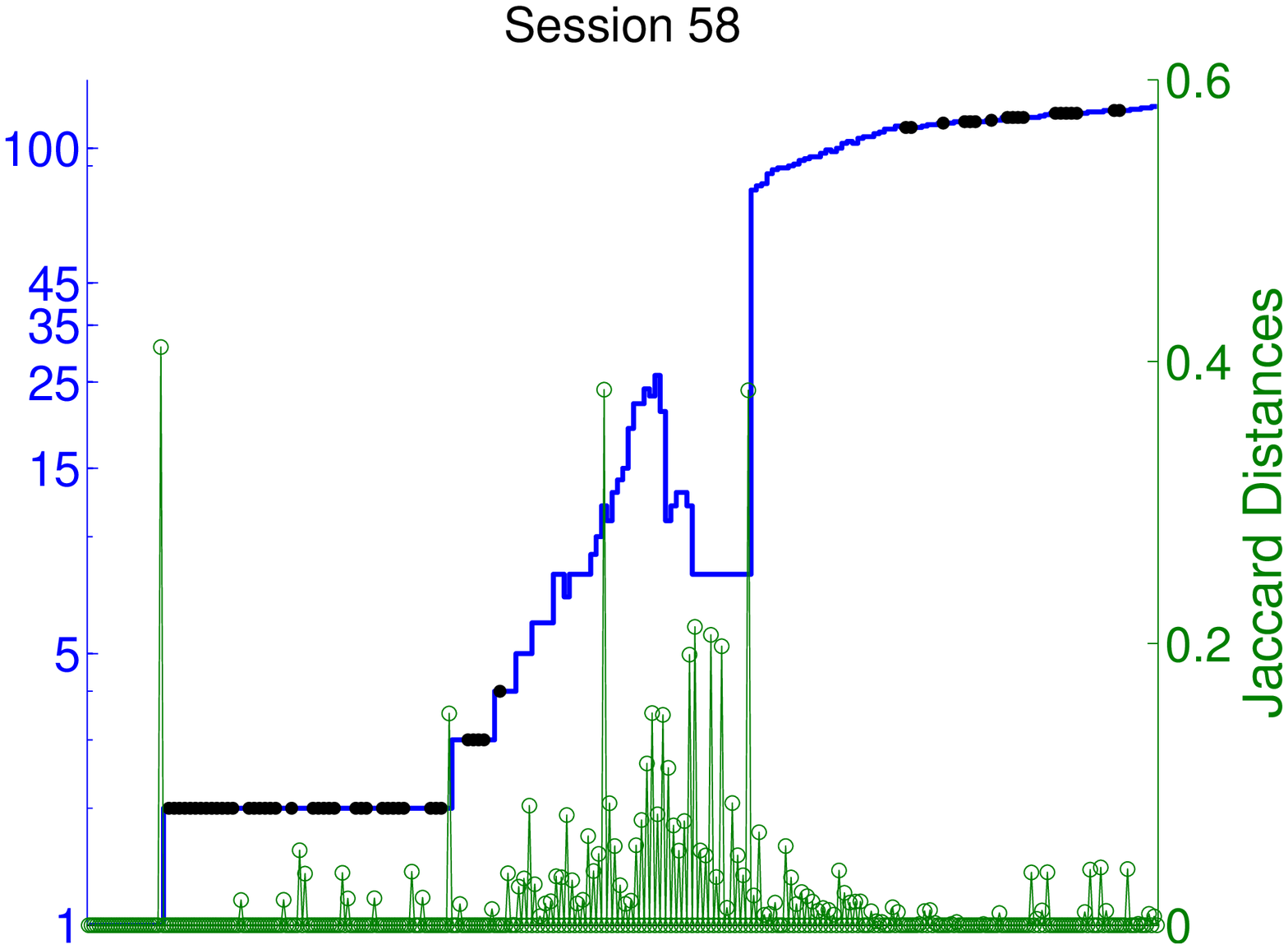}}
  \vspace*{-0.1in}
  \centerline{\begin{minipage}[c]{0.33\textwidth}\hspace*{-0.025in}\includegraphics[width=.925\textwidth]{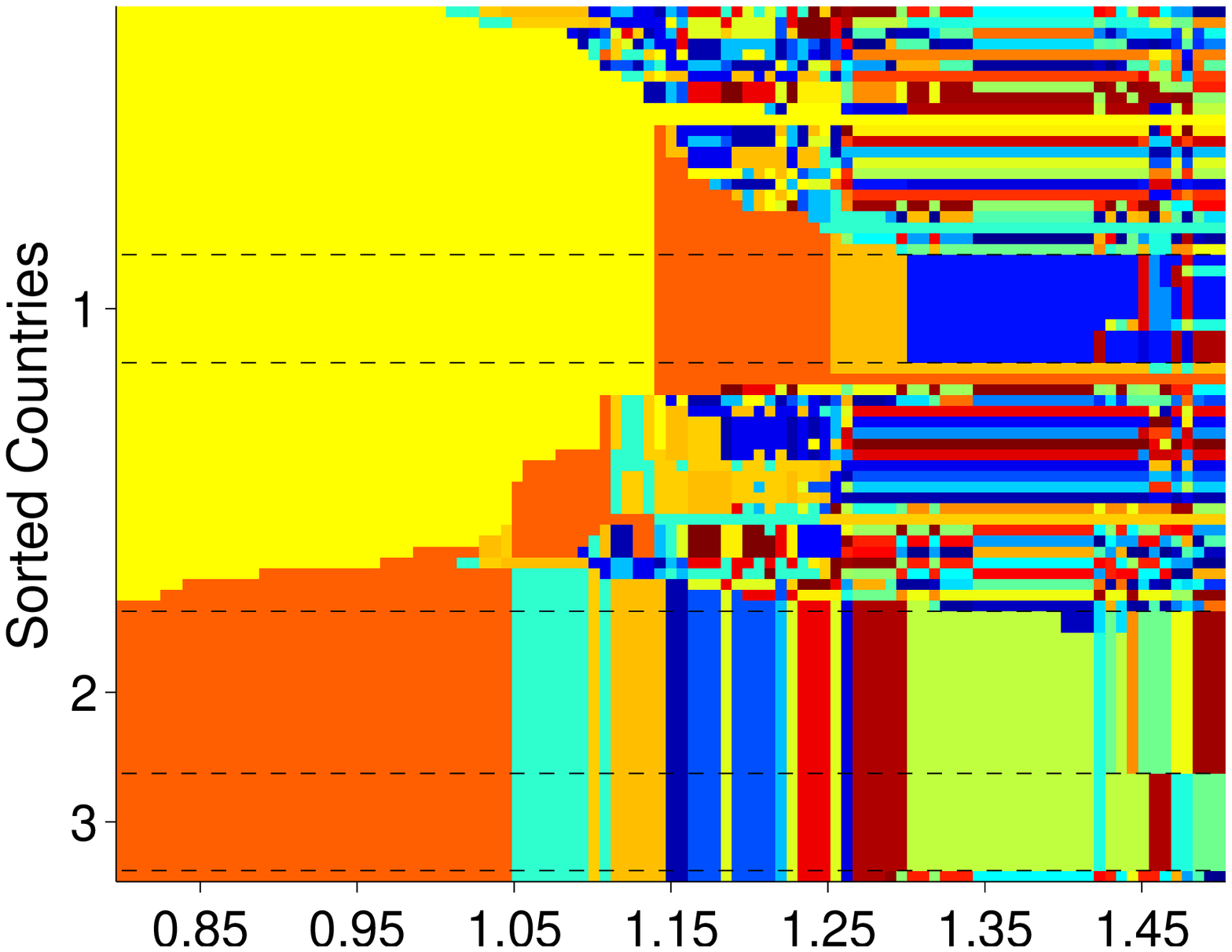}\end{minipage}\begin{minipage}[c]{0.33\textwidth}\vspace*{0.11in}\includegraphics[width=.89\textwidth,height=1.78in]{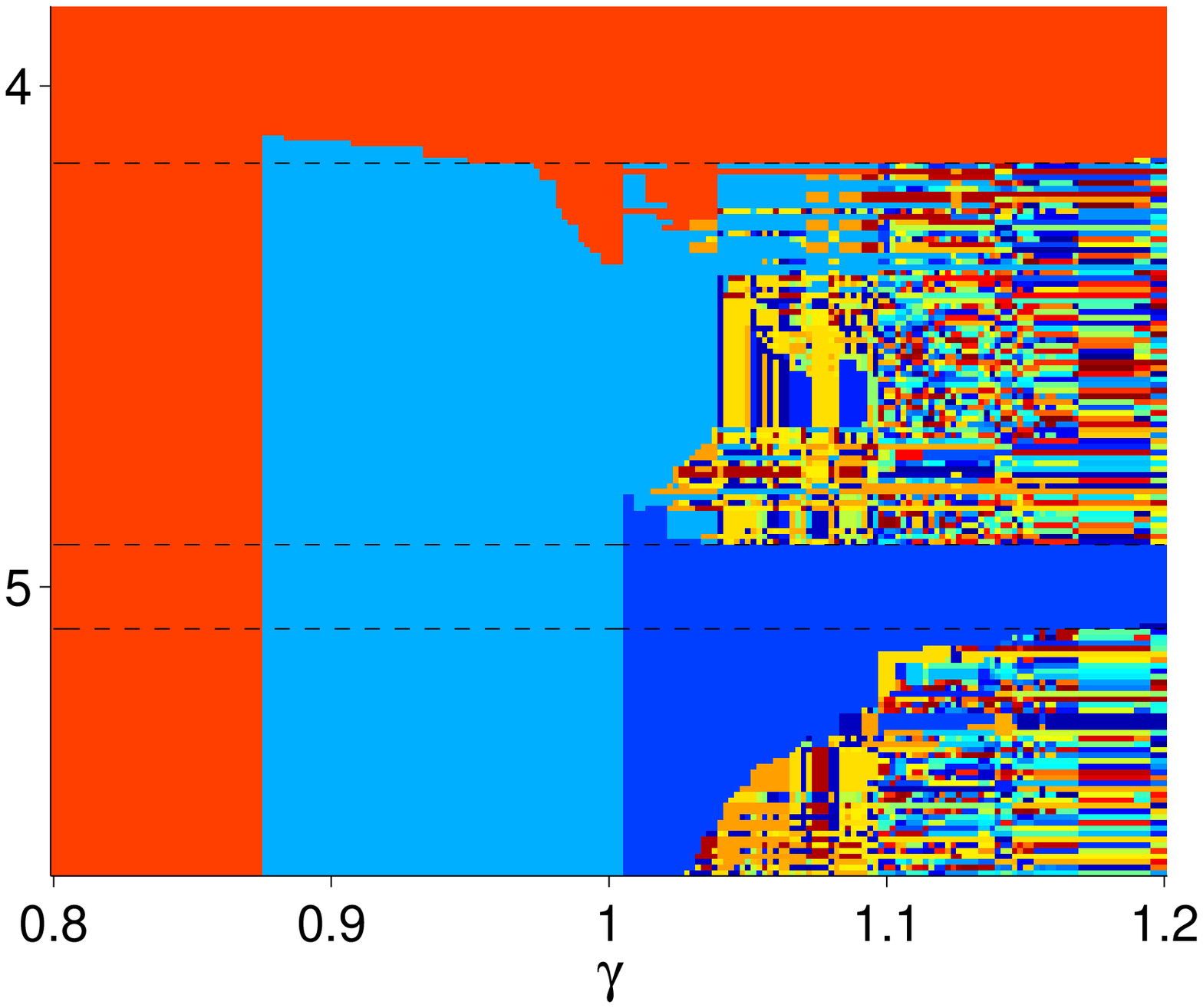}\end{minipage}\begin{minipage}[c]{0.33\textwidth}\hspace*{-0.175in}\includegraphics[width=.89\textwidth]{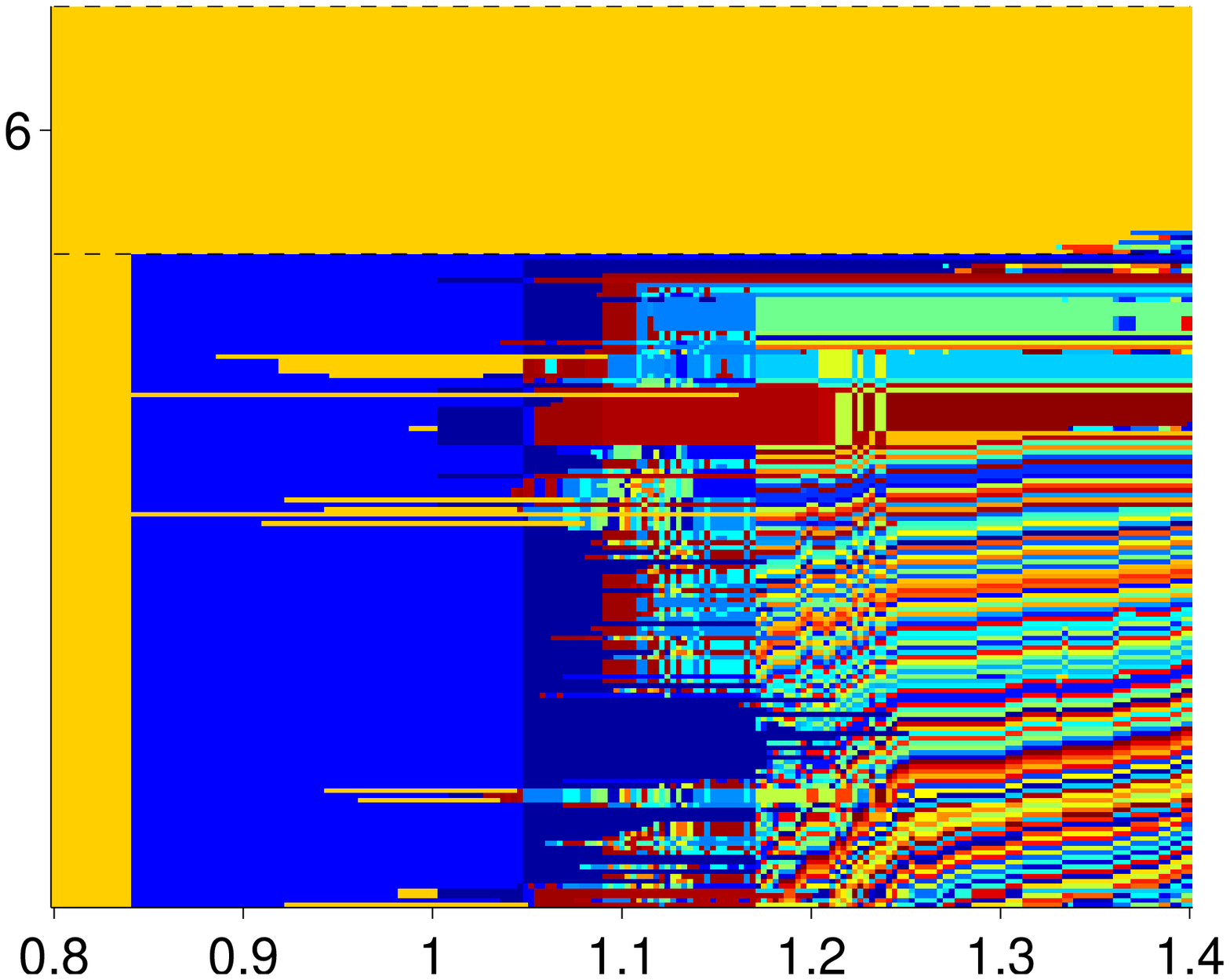}\end{minipage}}
  \caption{(Color) Results of community detection in UNGA Sessions 11 (left), 36 (middle), \& 58 (right) by optimizing the modularity in (\ref{Q1}) for the network of voting similarities. (Top) Number of communities ($N_c$; blue) that we obtained at each value of the resolution parameter $\gamma$ and the Jaccard distance (green) between partitions obtained at neighboring values of $\gamma$ (which differ by $\Delta\gamma \approx 0.007$, $0.002$, and $0.003$, respectively).
(Bottom) Community assignments at each value of $\gamma$; we use color to visualize each community, and we sort countries vertically by their community assignments across $\gamma$.  In Table \ref{subgroup}, we indicate the three groups of countries in Session 11 countries that are grouped together robustly: the Western core of communities who are commonly grouped together (1) and an Eastern community containing subgroups of Abstaining countries (2) and ``No"-Voting countries (3).   In Table \ref{36table}, we indicate the two groups of countries that appear robustly across this range of resolution parameter values for Session 36; given the identities of the countries, they can be labeled as a Western Community (4) and an Eastern Community (5).  In Table \ref{58table}, we similarly identify a robust group of countries as a ``North'' community (6).}
  \label{plots113658}
\end{figure*}

Because one has to be careful with modularity's resolution limit \cite{resolution}, we will examine these observations more carefully in single-session case studies by incorporating resolution parameters and the more sophisticated null models described above.  In examining the UNGA community structure over time, we consider three key eras: the early years of the Cold War (Sessions 1--25; \emph{1946--1970}), a transitional period (Sessions 26--45; \emph{1971--1990}), and the post-Cold War era (Sessions 46--63; \emph{1991--2008}).  
We consider one case study from each era---Sessions 11, 36, and 58---and start by optimizing modularity using the null model from equation (\ref{Q1}) over a large range of resolution parameter values $\gamma$, starting roughly from where the network begins to split up and ending when we obtain a set of communities that consist of individual countries. 
We seek regions of resolution parameters with identical or very similar community partitions, which then suggests that we have detected robust mesoscopic features of the network \cite{spinglass,fenn,reid,ourreview,santobig}.  The sizes of such regions can also potentially provide hints about the extent of such robustness. To find meaningful values of $\gamma$, we track changes between the numbers of communities obtained at neighboring parameter values and seek large regions (that is, plateaus) in which the network is partitioned into the same number of communities.  We also quantify changes between neighboring partitions using the Jaccard distance, although other quantities can also be used \cite{facebook,santobig}.  We show the Jaccard distance between neighboring values of $\gamma$ for Session 11 in the top left panel of Fig.~\ref{plots113658}.

In order to assist in the visualization of common associations, we sorted the UNGA member  countries using the community assignments in the network partitions.  In the bottom left panel of Fig.~\ref{plots113658}, we illustrate the communities that we obtained for Session 11.  We find a large region ($.8 \leq \gamma \lesssim 1$) of two-community partitions that differ only in the placement of five countries (Finland, Cambodia, Ethiopia, Iraq, and Japan).  The persistent orange community in this $\gamma$ range is the ``East'' group of countries, and the persistent yellow group of countries is the ``West".  Increasing the resolution parameter splits the Western community into several much smaller communities (including some consisting of individual countries), leaving a small core bloc (see Table~\ref{subgroup}) of countries stably placed together for $\gamma$ up to and including the observed plateau at $\gamma \in (1.353,1.388)$. Countries in the core group inside the Western bloc agree more with each other than they do with the rest of the Western bloc.  Meanwhile, the Eastern community splits into two subgroups (Table~\ref{subgroup} lists the country names in each of these groups): a group voting ``no'' on 43\% of the resolutions (and abstaining on 10\% of them) and another voting ``abstain'' on 34\% of the resolutions (and voting ``no" on 9\% of them).  The group of countries that votes ``no" also has fewer agreements with the West than the countries in the ``abstain'' group.

Although the community sizes drop off faster with resolution parameter in Session 36 and Session 58 (see, respectively, the middle and right parts of Fig.~\ref{plots113658}) than they do in Session 11, we again find a dominant two-community region ($0.8 \leq \gamma \lesssim 1$) for both sessions.  For Session 36, we list the countries in the Eastern bloc and Western bloc communities in Table~\ref{36table}.  Session 36 also has a small plateau with three communities for $\gamma$ just above $1$.  In Session 58, the dominant two-community split divides countries into North and South groups (as indicated previously at $\gamma=1$), and we similarly find a small plateau in which the network is partitioned into three communities for $\gamma$ just above $1$.  After these communities split up with a further increase of $\gamma$, we do not find any plateaus of reasonable size that correspond to network partitions with more communities, and only one group of countries appears to remain robust across this $\gamma$ range (the North group listed in Table~\ref{58table}).


\section{Networks of Vote Agreements and Disagreements} \label{vote}

\begin{figure*}
\centerline{\begin{minipage}[c]{.38\textwidth}
\includegraphics[width=\textwidth]{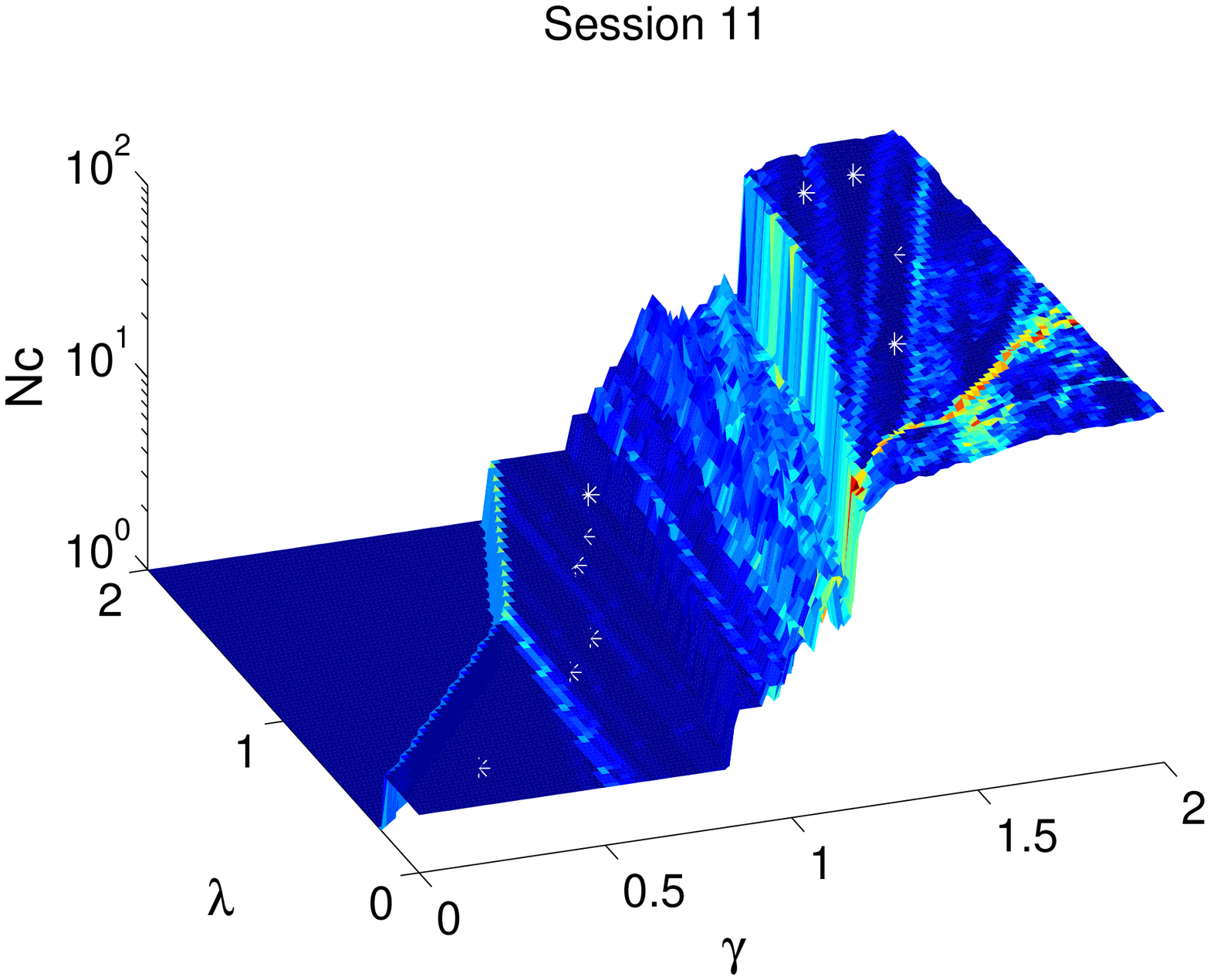}
\end{minipage}\hspace*{-0.25in}
\begin{minipage}[c]{.28\textwidth}
\begin{tabular}{r | r | r}
Index & $\gamma$ & $\lambda$\\
\hline
1&        .24   &  .22 \\
2&        .66   &  .70 \\
3&        .78   &  .88 \\
4&        .90   &1.32 \\
5&        .98   &1.48 \\
6&      1.08   &1.72 \\
7&      1.68   &1.78 \\
8&      1.82   &1.80 \\
9&      1.74   &1.28 \\
10&    1.56   &  .78 \\
\end{tabular}
\end{minipage}
\begin{minipage}[c]{.3\textwidth}
\includegraphics[width=\textwidth]{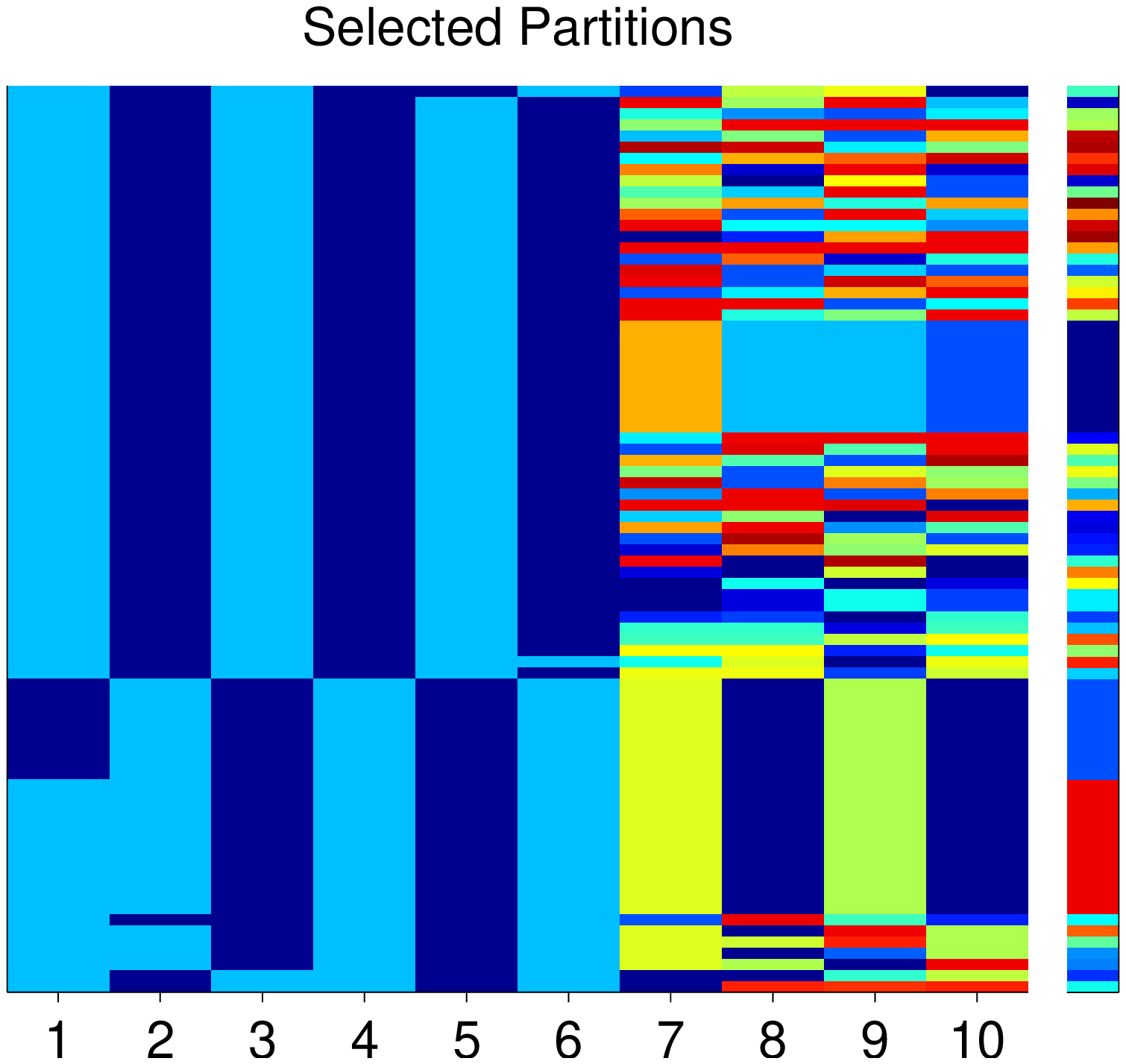}
\end{minipage}}

\centerline{\begin{minipage}[c]{.38\textwidth}
\includegraphics[width=\textwidth]{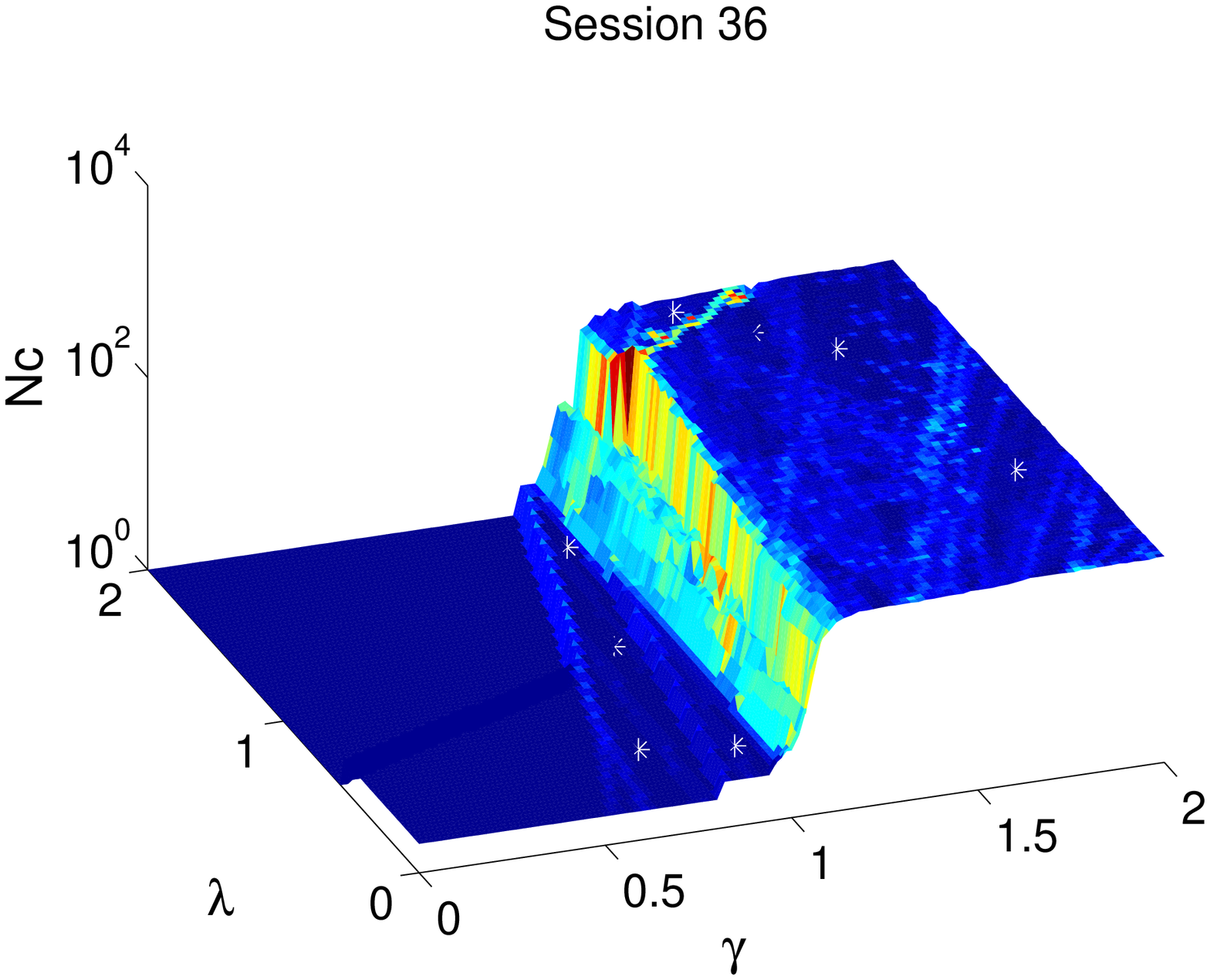}
\end{minipage}\hspace*{-0.25in}
\begin{minipage}[c]{.28\textwidth}
\begin{tabular}{r | r | r}
Index & $\gamma$ & $\lambda$\\
\hline
1&  .72    & .36    \\
2&  .92    & .20    \\
3&  .88    & .98    \\
4&  .98    & 1.60  \\
5&  1.38  & 1.92  \\
6&  1.52  & 1.72  \\
7&  1.68  & 1.54  \\
~8&  1.84  & .66  \\
\end{tabular}
\end{minipage}
\begin{minipage}[c]{.3\textwidth}
\hspace*{-0.08\textwidth}
\includegraphics[width=1.06\textwidth]{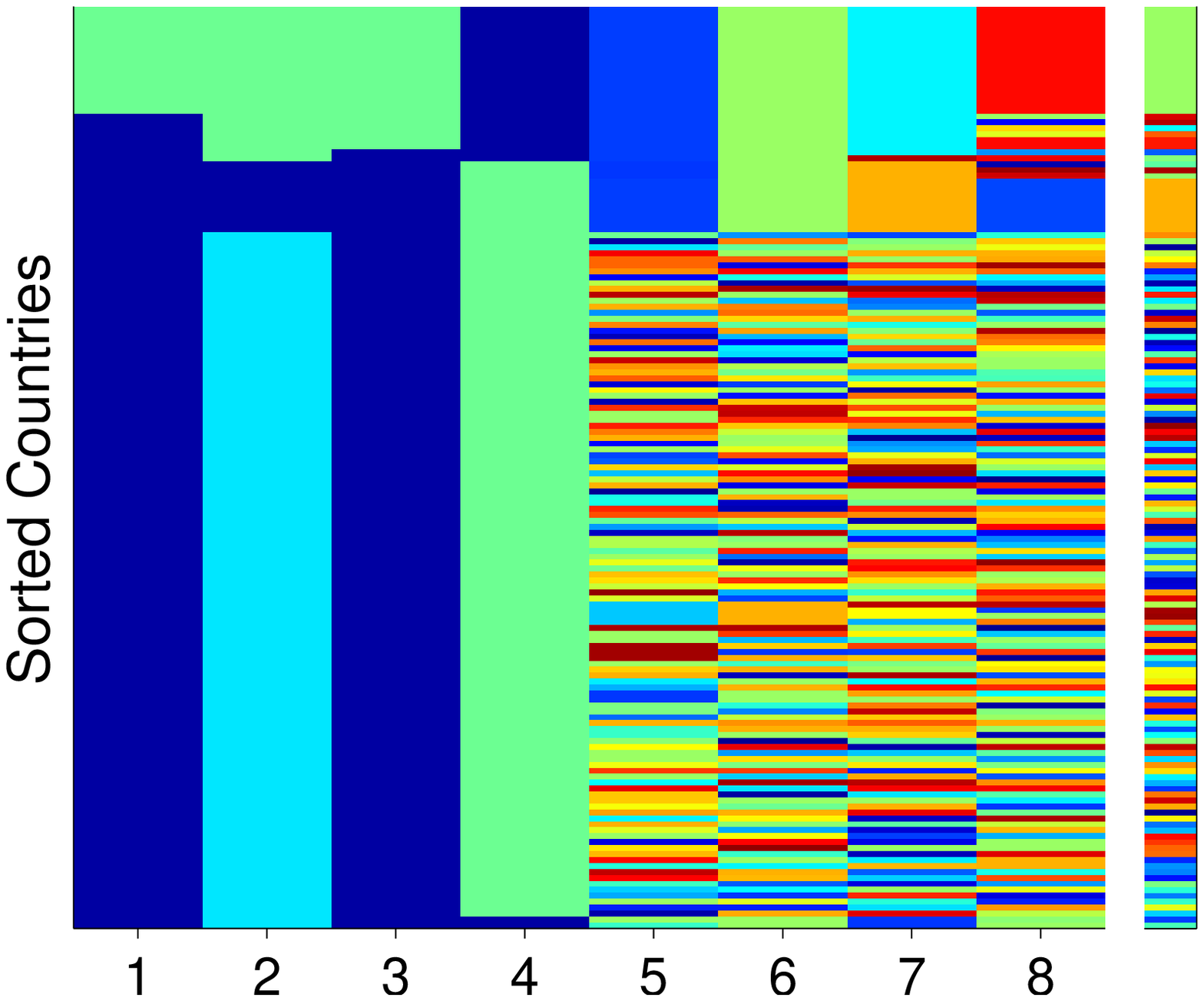}
\end{minipage}}

\centerline{\begin{minipage}[c]{.38\textwidth}
\includegraphics[width=\textwidth]{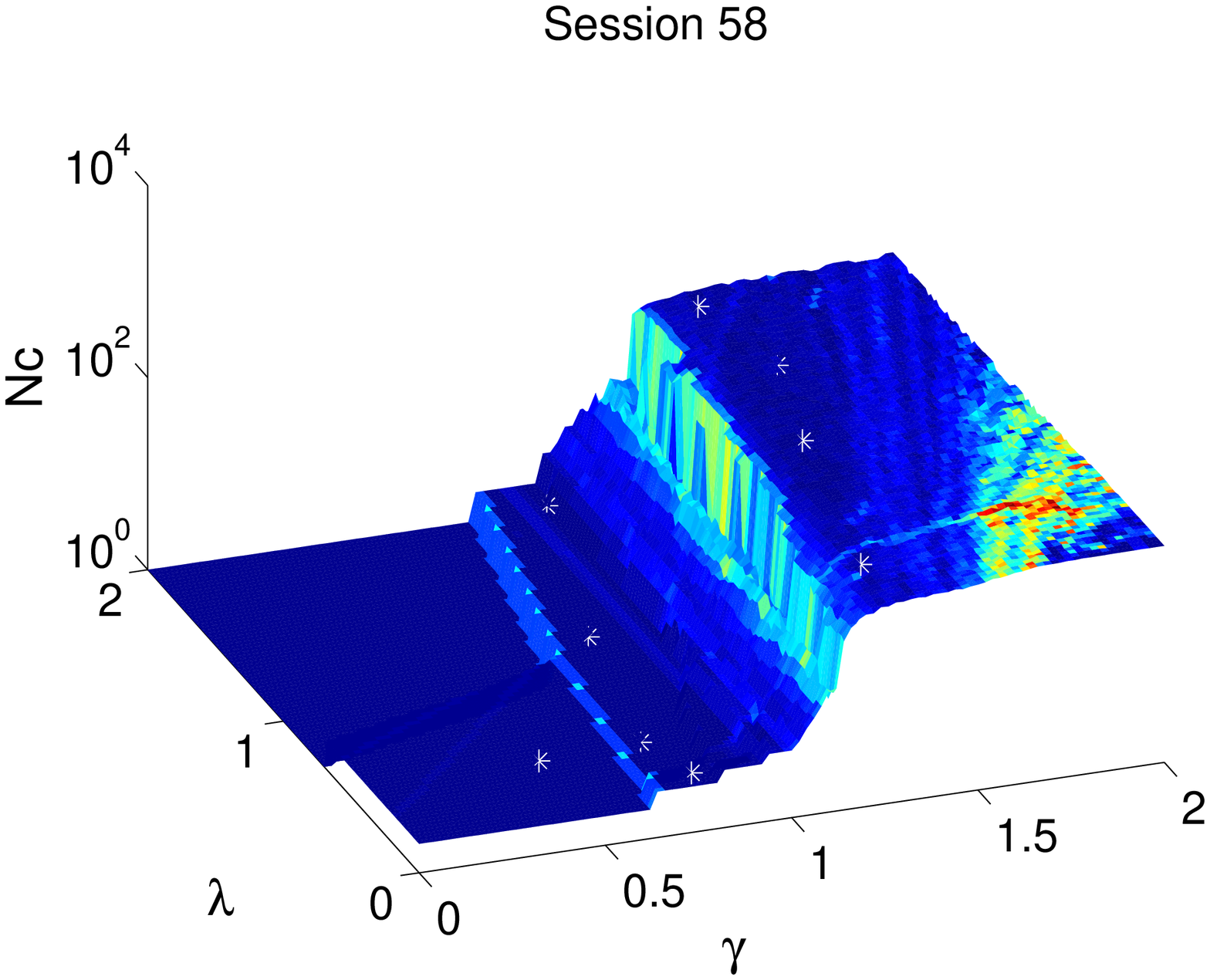}
\end{minipage}\hspace*{-0.25in}
\begin{minipage}[c]{.28\textwidth}
\begin{tabular}{r | r | r}
Index & $\gamma$ & $\lambda$\\
\hline
 1 & .46   &    .38  \\
 2 & .76   &    .08  \\
 3 & .74   &    .40  \\
 4 & .84   &  1.06  \\
 5 & 1.02 &  1.86  \\
 6 & 1.44 &  1.90  \\
 7 & 1.50 &  1.48  \\
 8 & 1.40 &  1.02  \\
 ~9 & 1.28 &  .26  \\
\end{tabular}
\end{minipage}
\begin{minipage}[c]{.3\textwidth}
\includegraphics[width=\textwidth]{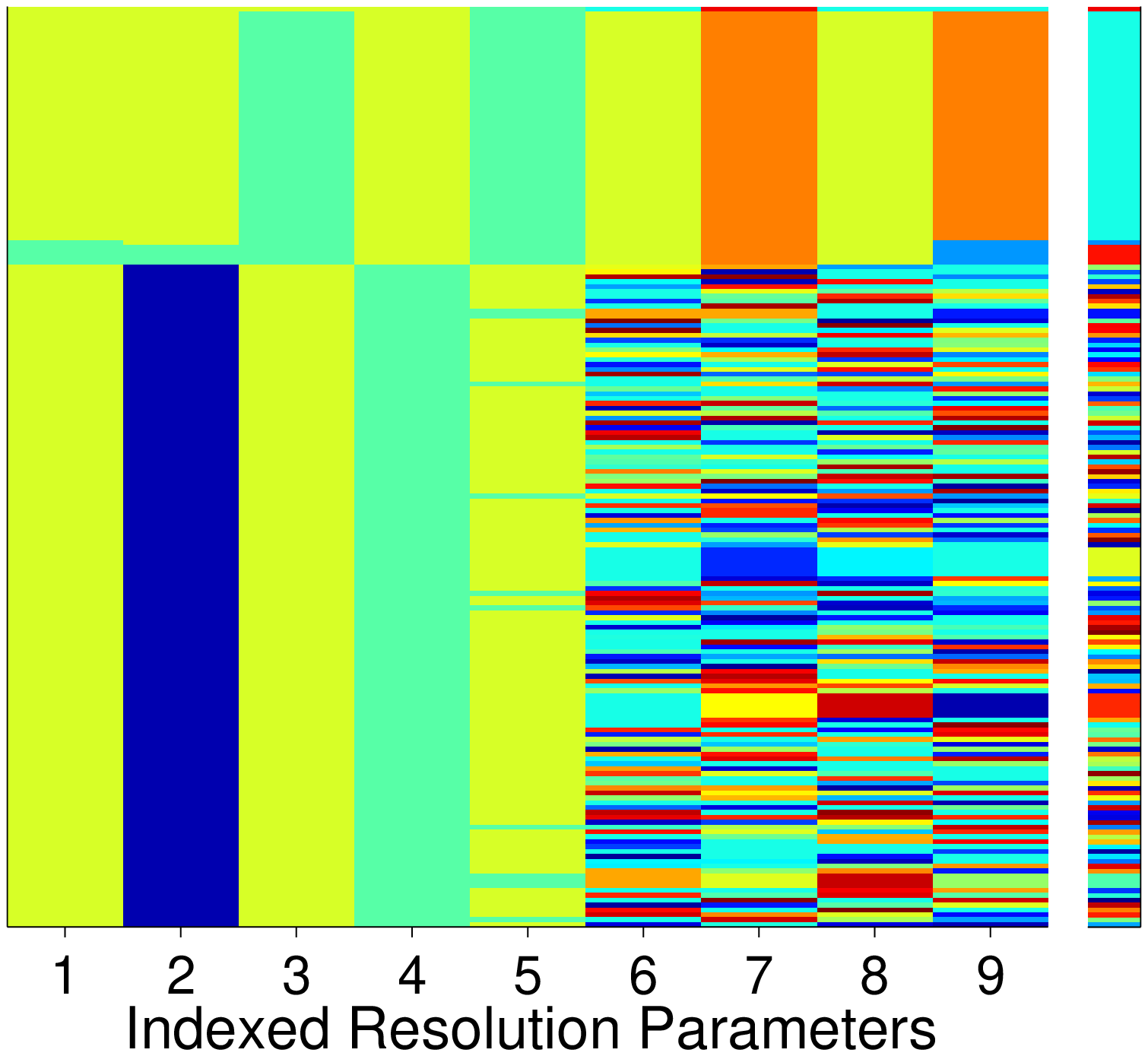}
\end{minipage}}
\caption{(Color) Exploration of the resolution parameter space for the signed, unipartite network of countries in Sessions 11 (top), 36 (middle), and 58 (bottom) of the UNGA.  (Left) Number of communities identified at each pair of resolution parameter values, which we have color-coded according to the mean Jaccard distance between the partition and its four nearest neighbors in the space of resolution parameters. (Center) Resolution parameter values in different stable regions (selected by hand) of the resolution parameter space.  (Right) Color-coded visualization of the communities that we obtained at each indexed point in resolution parameter space, with an additional column (far right) that color-codes the blocks of countries and resolutions that are grouped together robustly at all of the indexed partitions.}
\label{fig:signed}
\end{figure*}

In this section, we study the signed unipartite networks that we obtain by treating the expected number of positive and negative edges in the network separately.  As indicated in Section~\ref{commdetect}, this yields a null model with two terms and a resolution parameter ($\gamma$ and $\lambda$) for each of them.  Using the network of agreements and disagreements described in Section~\ref{data}, we sweep over different values of the two resolution parameters and plot surfaces for the numbers of communities in Fig.~\ref{fig:signed}. In these plots, we have color-coded the surface by the mean Jaccard distances between each partition and its nearest neighbors in resolution parameter space. That is, given the square grid of sampled resolution parameter values that we explore, each partition is compared with its four nearest neighbors.

In this two-dimensional parameter space, one can no longer easily visualize all of the community assignments at each resolution parameter value.  Because we seek robust community assignments, we avoid parameter values at which the number of communities (the height in the left panels of Fig.~\ref{fig:signed}) change rapidly.  In order to ensure similarities in nearby partitions (cf.\ the weaker restriction of the same numbers of communities), we calculate the Jaccard distances between the partitions obtained at nearest-neighbor points on a square grid in the ($\gamma$,$\lambda$) plane, with both dimensions discretized by 101 uniform points, $(\gamma,\lambda)\in[0,2]$.  We have color-coded the left panels of Fig.~\ref{fig:signed} at each grid point by the mean Jaccard distance between that partition and its four nearest neighbors.  We then selected points on this grid by hand to correspond to community assignments that persist over a range of resolution parameter values (we indicate these in the figure).  For each of our three case studies (Sessions 11, 36, and 58), we show the number of communities and mean Jaccard distance at the selected points in resolution parameter space.  In Fig.~\ref{fig:signed}, we have also tabulated the resolution parameter values and the partitions that we found at the selected points in parameter space.  In agreement with the results in Section \ref{agree}, we recover the dominate two-way split in the UNGA that spans a large portion of the parameter space.  Again as before, we find some smaller robust groups at multiple regions in parameter space.  Observe in Sessions 11 and 58 (and to a smaller extent in Session 36) that some communities of moderate and even large size persist robustly even when the other countries are partitioned by themselves or as members of tiny communities.  We identify these countries for Session 11, 36, and 58 in Tables~\ref{11signedTable}, \ref{36signedTable}, and~\ref{58signedTable}, respectively.


\section{Bipartite Voting Networks with Positive and Negative Edges} \label{bip}

In this section, we use the signed bipartite modularity from Section~\ref{bip} to study networks of yes and no votes, which we represent using positive and negative edges between UNGA countries and the resolutions on which they voted.  (We do not include abstentions in this network.)  As before, we seek robust communities by exploring the two-dimensional space of resolution parameter values and examining the number of communities and mean Jaccard distances between partitions obtained at nearest-neighbor resolution parameter values on a uniform grid.  The Jaccard distances that we used for the partitions of the bipartite voting network were calculated with respect to the full bipartite network (rather than restricted to the partitions of countries).

\begin{figure*}
\centerline{\begin{minipage}[c]{.38\textwidth}\includegraphics[width=\textwidth]{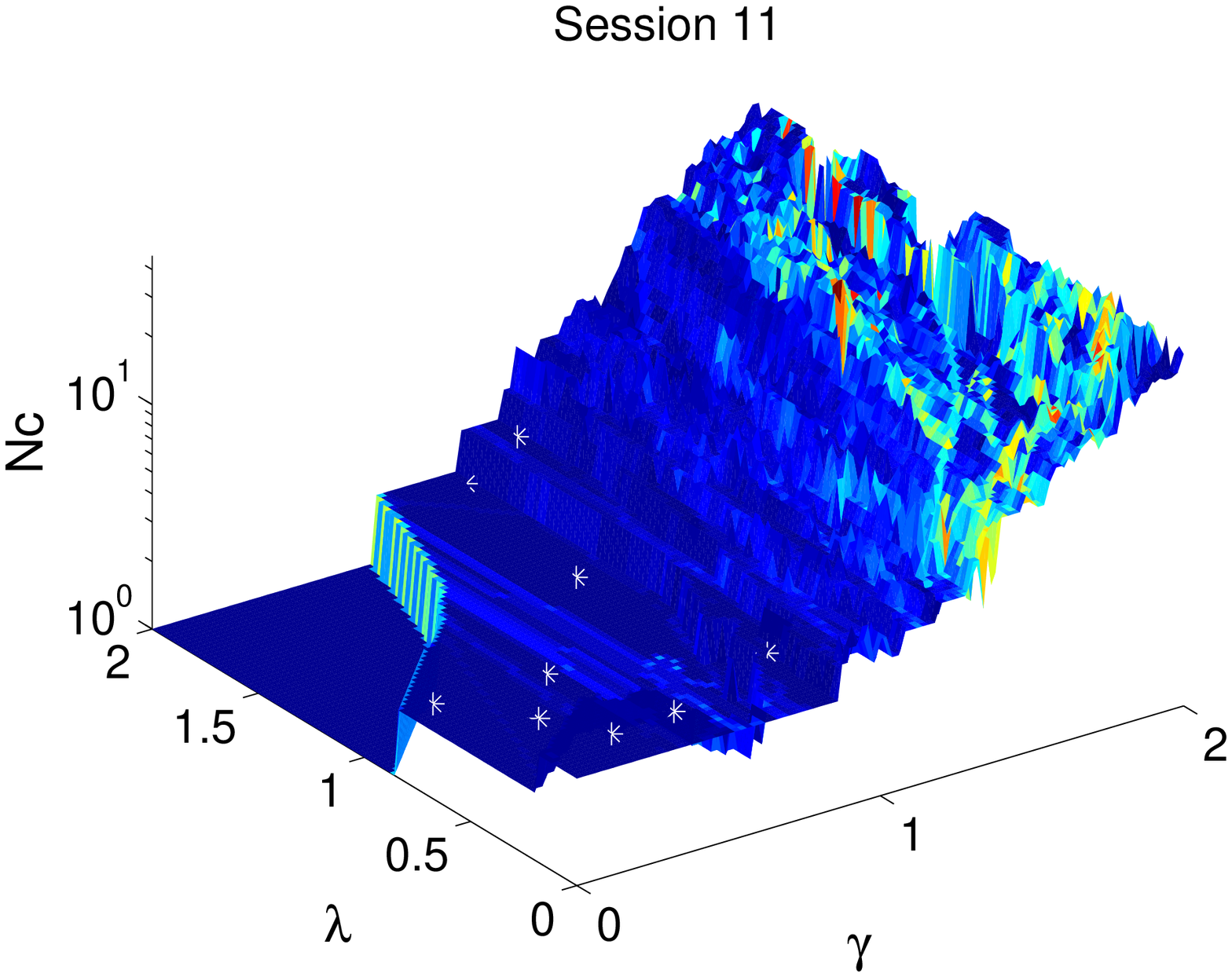}
\end{minipage}\hspace*{-0.25in}
\begin{minipage}[c]{.28\textwidth}
\begin{tabular}{r | r | r}
Index & $\gamma$ & $\lambda$\\
\hline
 1& 1.12 &   1.88 \\
 2&   .98 &    1.92 \\
 3&   .88 &   1.26 \\
 4&   .88 &    .36 \\
 5&   .46 &    .80 \\
 6&   .46 &    .20 \\
 7&   .24 &    .18 \\
 8&   .28 &    .58 \\
 9&   .10 &    .82 \\
\end{tabular}
\end{minipage}
\begin{minipage}[c]{.3\textwidth}
\includegraphics[width=\textwidth]{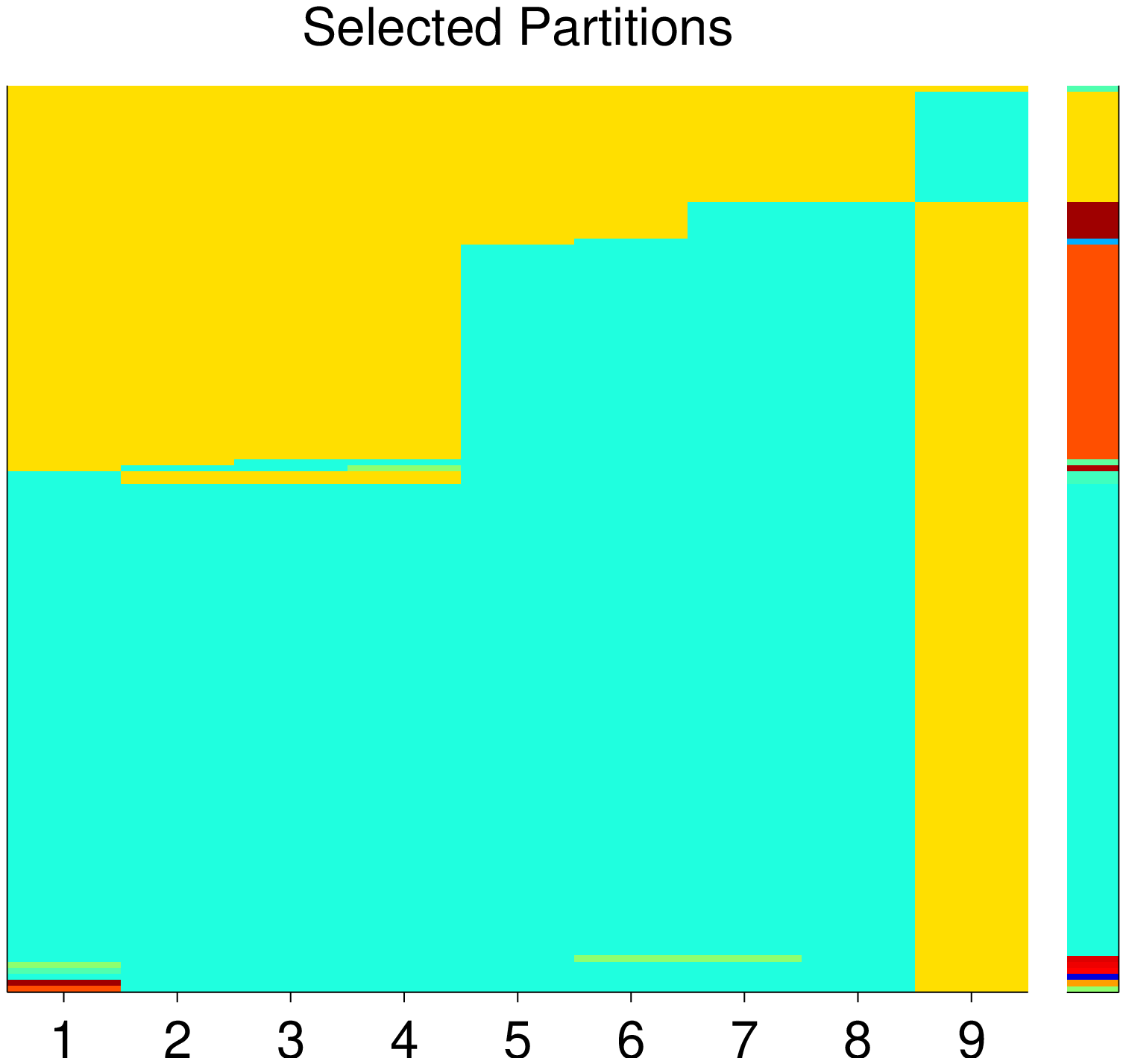}
\end{minipage}}

\centerline{\begin{minipage}[c]{.38\textwidth}
\includegraphics[width=\textwidth]{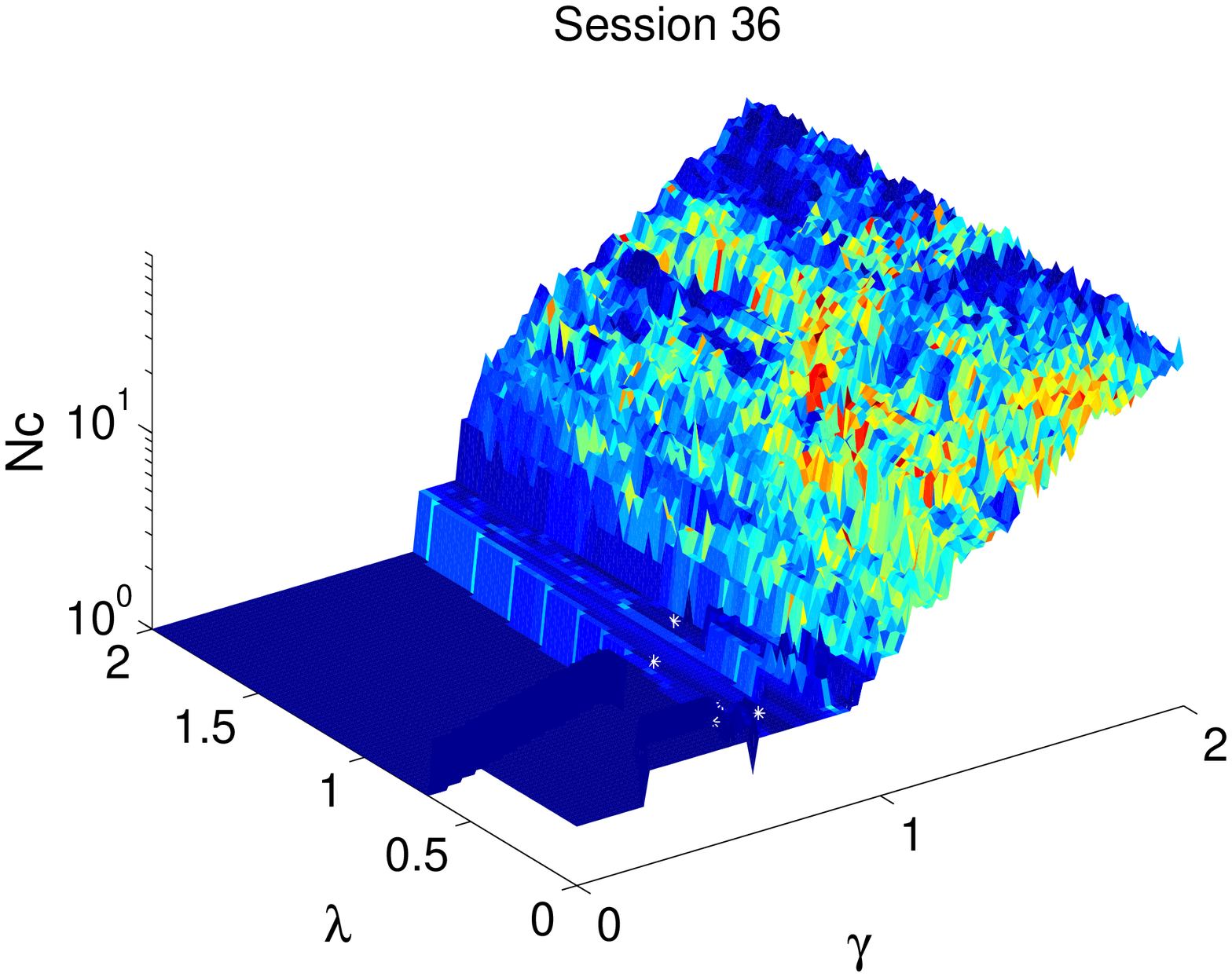}
\end{minipage}\hspace*{-0.25in}
\begin{minipage}[c]{.28\textwidth}
\begin{tabular}{r | r | r}
Index & $\gamma$ & $\lambda$\\
\hline
1&    .558   &  .156 \\
2&    .558   &  .408 \\
3&    .774   &  .420 \\
4&    .774   &  .744 \\
5&    .930   &  .060 \\
6&    .966   &1.176 \\
7&    .966   &  .924 \\
8&    .690   &  .132 \\
\end{tabular}
\end{minipage}
\begin{minipage}[c]{.3\textwidth}
\hspace*{-0.08\textwidth}
\includegraphics[width=1.06\textwidth]{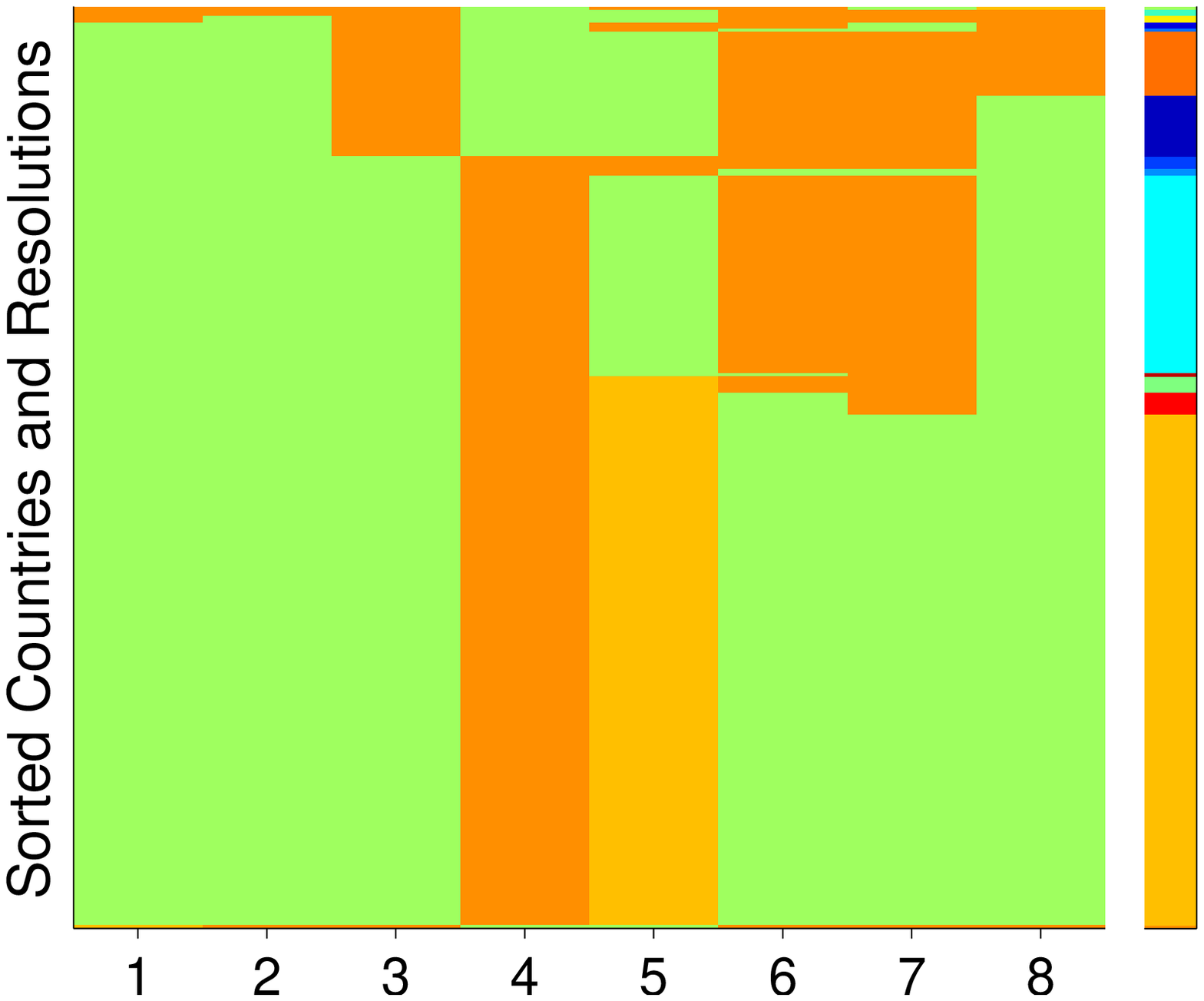}
\end{minipage}}

\centerline{\begin{minipage}[c]{.38\textwidth}
\includegraphics[width=\textwidth]{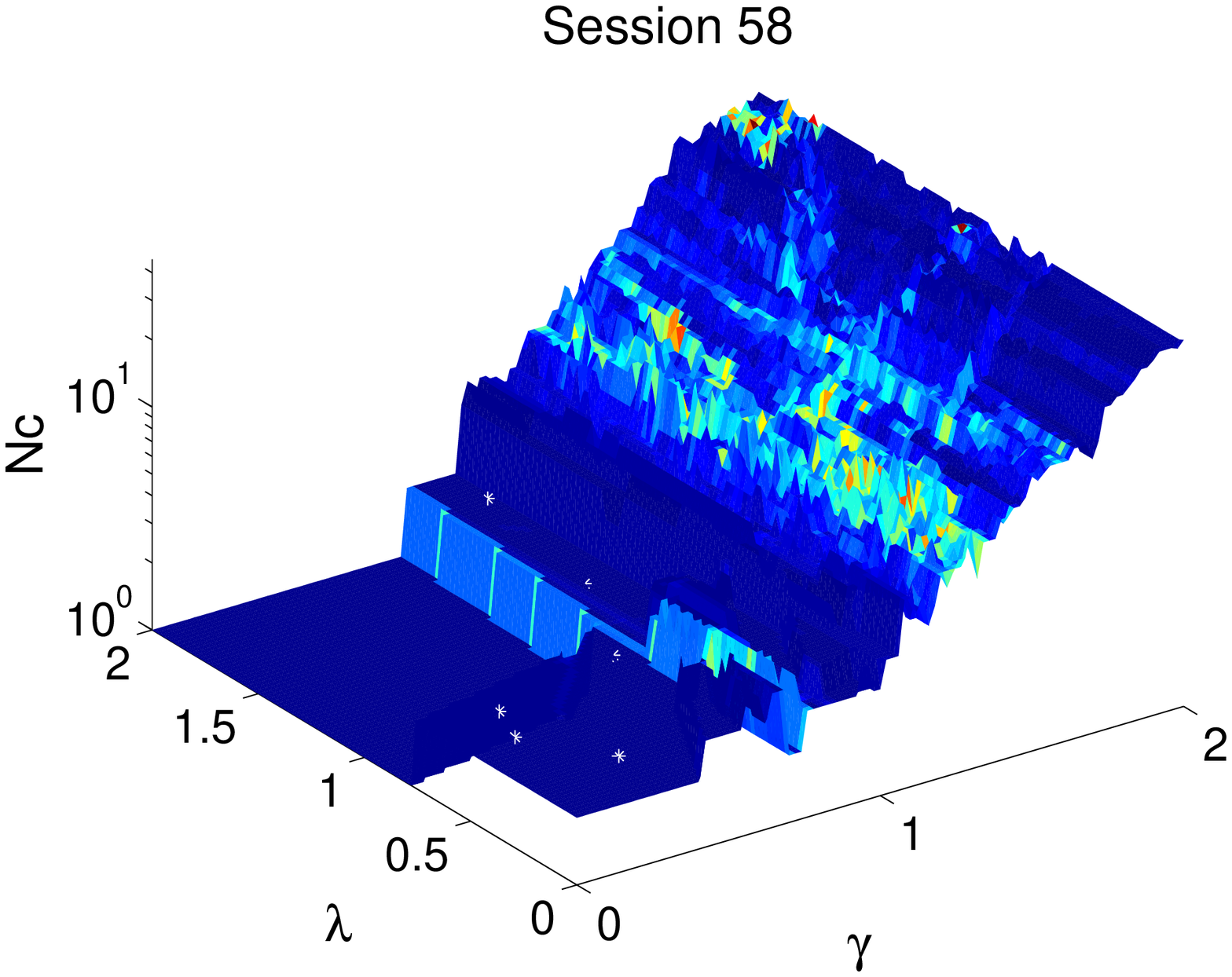}
\end{minipage}\hspace*{-0.25in}
\begin{minipage}[c]{.28\textwidth}
\begin{tabular}{r | r | r}
Index & $\gamma$ & $\lambda$\\
\hline
1&    .88&  1.22 \\
2&    .68&    .80 \\
3&    .32&    .26 \\
4&    .16&    .52 \\
5&    .22&    .68 \\
6&    .90&    .52 \\
7&    .98&  1.82 \\
\end{tabular}
\end{minipage}
\begin{minipage}[c]{.3\textwidth}
\includegraphics[width=\textwidth]{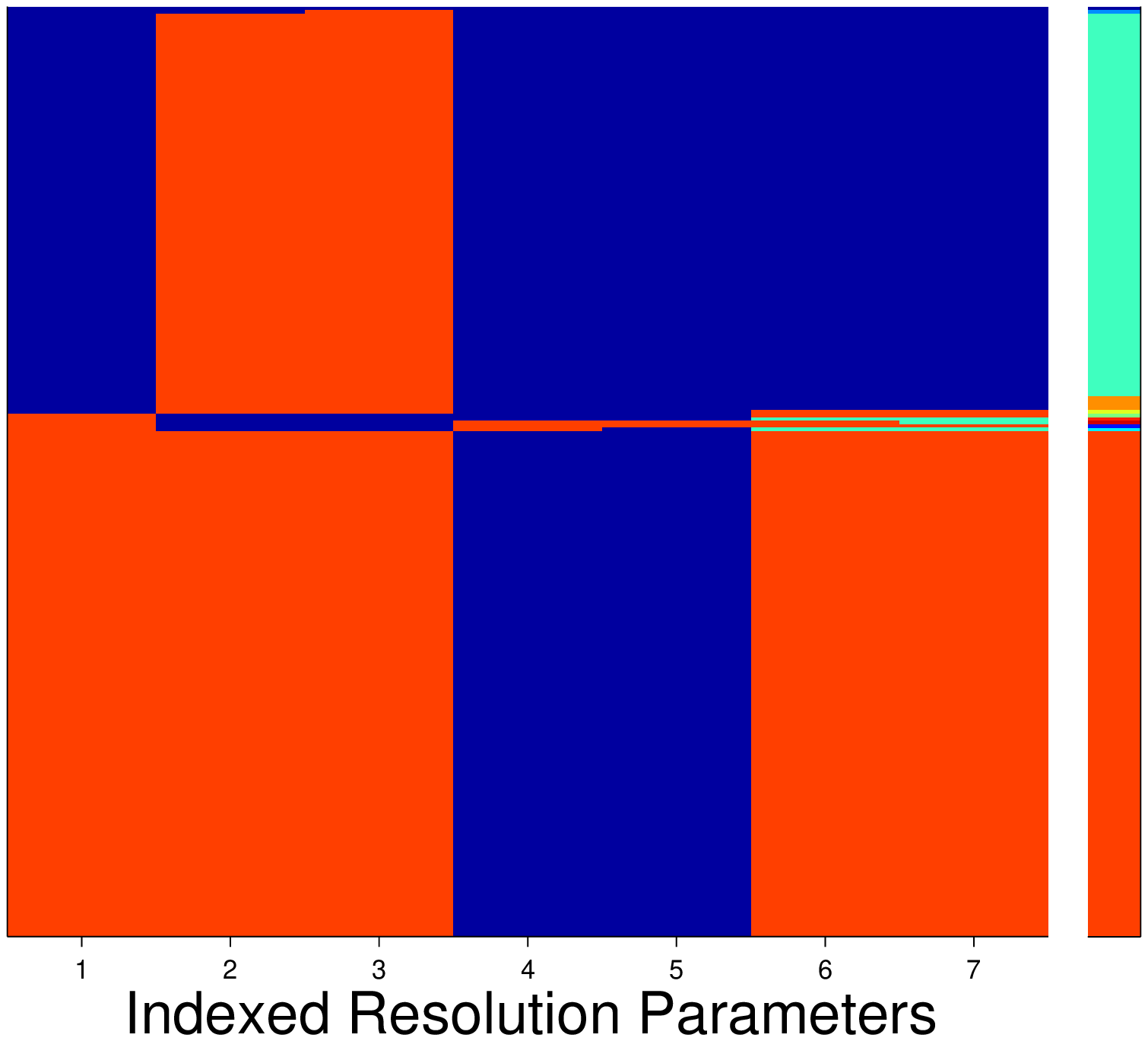}
\end{minipage}}
\caption{(Color) Exploration of the resolution parameter space for the signed, bipartite network of countries and resolutions in UNGA Sessions 11 (top), 36 (middle), and 58 (bottom). (Left) Number of communities that we identified for each value of the pair of resolution parameters.  We have color-coded these plots using the mean Jaccard distance (which we computed for the full bipartite network) between the partition and its four nearest neighbors in resolution parameter space. (Center) Resolution parameter values for different stable regions in the resolution parameter space (selected by hand).  (Right) Color-coded visualization of communities that we obtained at each indexed point in resolution parameter space.  The far right column color-codes the blocks of countries that are grouped together robustly at all of the indexed partitions.
}
\label{fig:bipart}
\end{figure*}

We show the results of our numerical exploration of the signed bipartite network representations in Fig.~\ref{fig:bipart}.  A major difference between our results for these networks and those of the previous sections is that community detection on the bipartite network of votes additionally includes UNGA resolutions in the groups with the countries that predominantly supported such families of resolutions.  We find that each of the three case-study sessions contains numerous resolution parameter values, identified in Fig.~\ref{fig:bipart}, in which there are large, robust communities (as we also saw with the other network formulations). Session 11 contains resolution parameter values in which one of these two robust communities is much larger than the other one. Sessions 36 and 58 both contain regions in which there is one dominant community that contains almost every country.  In contrast to the other network formulations, we note there do not appear to be any robust plateau regions in the $\gamma>1$ values in Fig.~\ref{fig:bipart} (cf.\ the results illustrated in Figs.~\ref{plots113658} and \ref{fig:signed}). Instead, each of the robust partitions identified in Fig.~\ref{fig:bipart} consists of two large groups with only a small number of countries assigned to additional groups.  While these partitions differ from one another across the resolution parameter landscape, many of the community assignments remain the same across these selected points, with the right edge of the figure identifying groups of countries that are placed together at all of the selected resolution parameter values.  In Tables~\ref{11bipartTable}, \ref{36bipartTable}, and~\ref{58bipartTable}, we list for Sessions 11, 36, and 58, respectively, the countries that appear robustly in the larger groups identified in the far right edge of Fig.~\ref{fig:bipart}.


\section{Conclusions and Discussion}\label{conc}

In this paper, we have studied the community structure of networks formed by voting on resolutions in individual sessions of the United Nations General Assembly.  The UNGA voting record provides a fascinating example of a very general problem: How can one use network methods such as community detection to examine data such as voting records?  Accordingly, our focus is not on attempting a sociological or political study of the UNGA but rather on using it as an interesting and potentially valuable example to consider different network frameworks that are each seemingly reasonable and then comparing the results that one can subsequently obtain from them.  To do this, we constructed networks from the UNGA voting records of sixty-three separate sessions between 1946 and 2008 in three different ways: (1) by considering voting similarities as weighted unipartite networks in the usual manner, (2) by considering voting similarities as weighted unipartite networks in a manner that takes the large extent of agreement of the General Assembly into account, and (3) as signed bipartite networks in which countries are connected to resolutions.  For each formulation, we detected communities by optimizing network modularity using an appropriate null model.  In optimizing a quality function such as modularity, the consideration of multiple resolution parameters enables one to examine different ``background" levels of agreement between nations.

\begin{figure*}
\centering

\hspace*{0.01in}
\includegraphics[width=.39\textwidth]{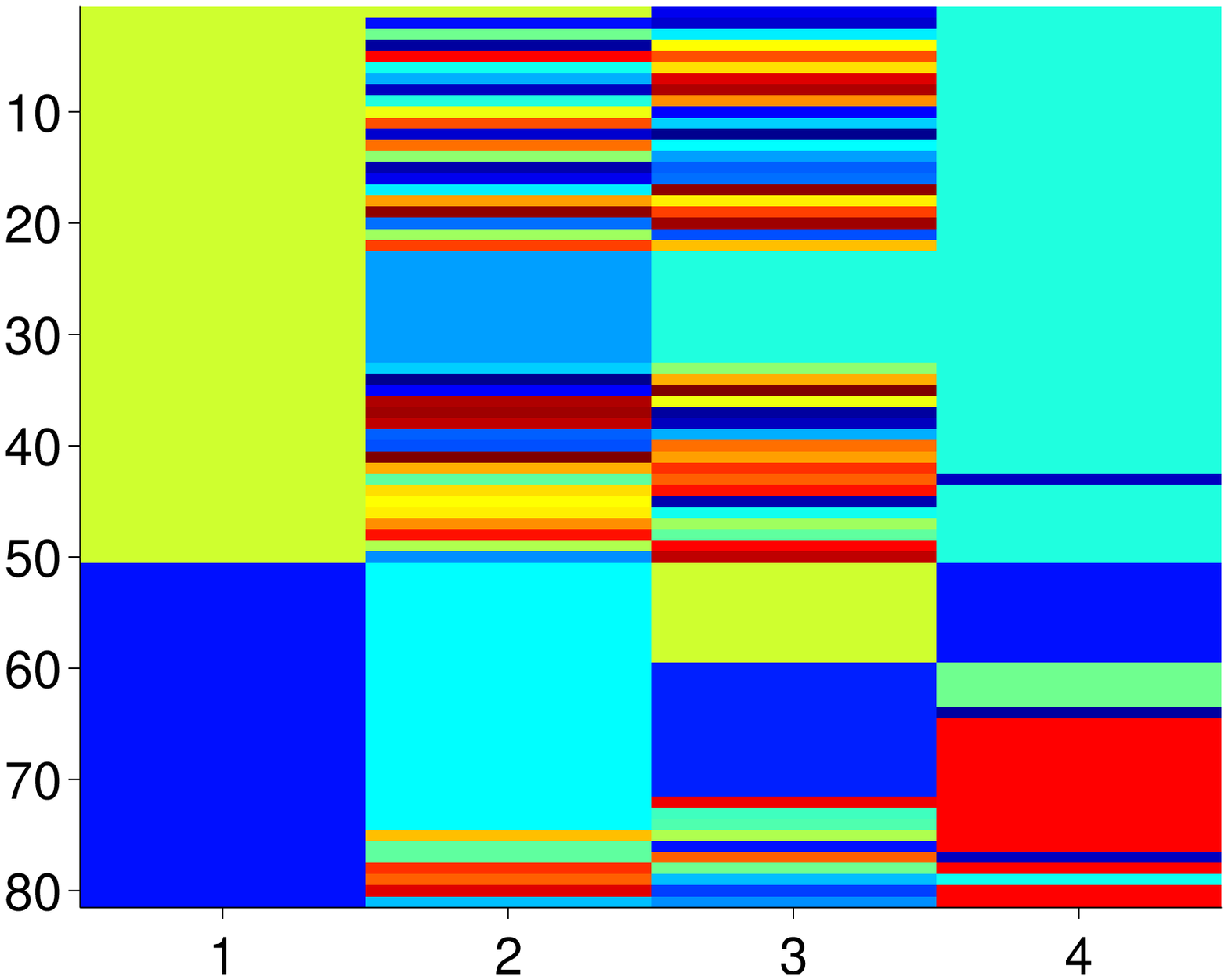}\\

\includegraphics[width=.4\textwidth]{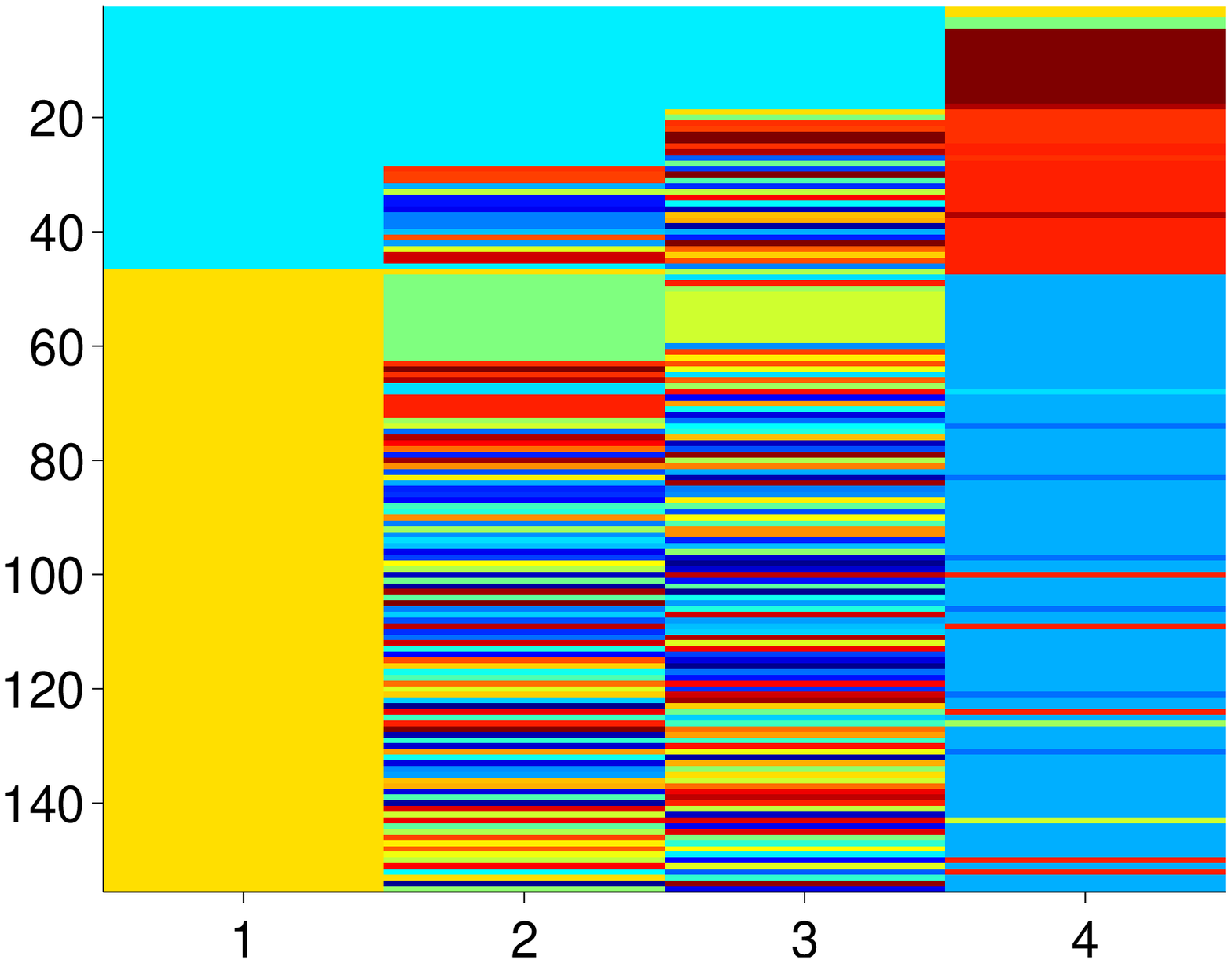}\\

\includegraphics[width=.4\textwidth]{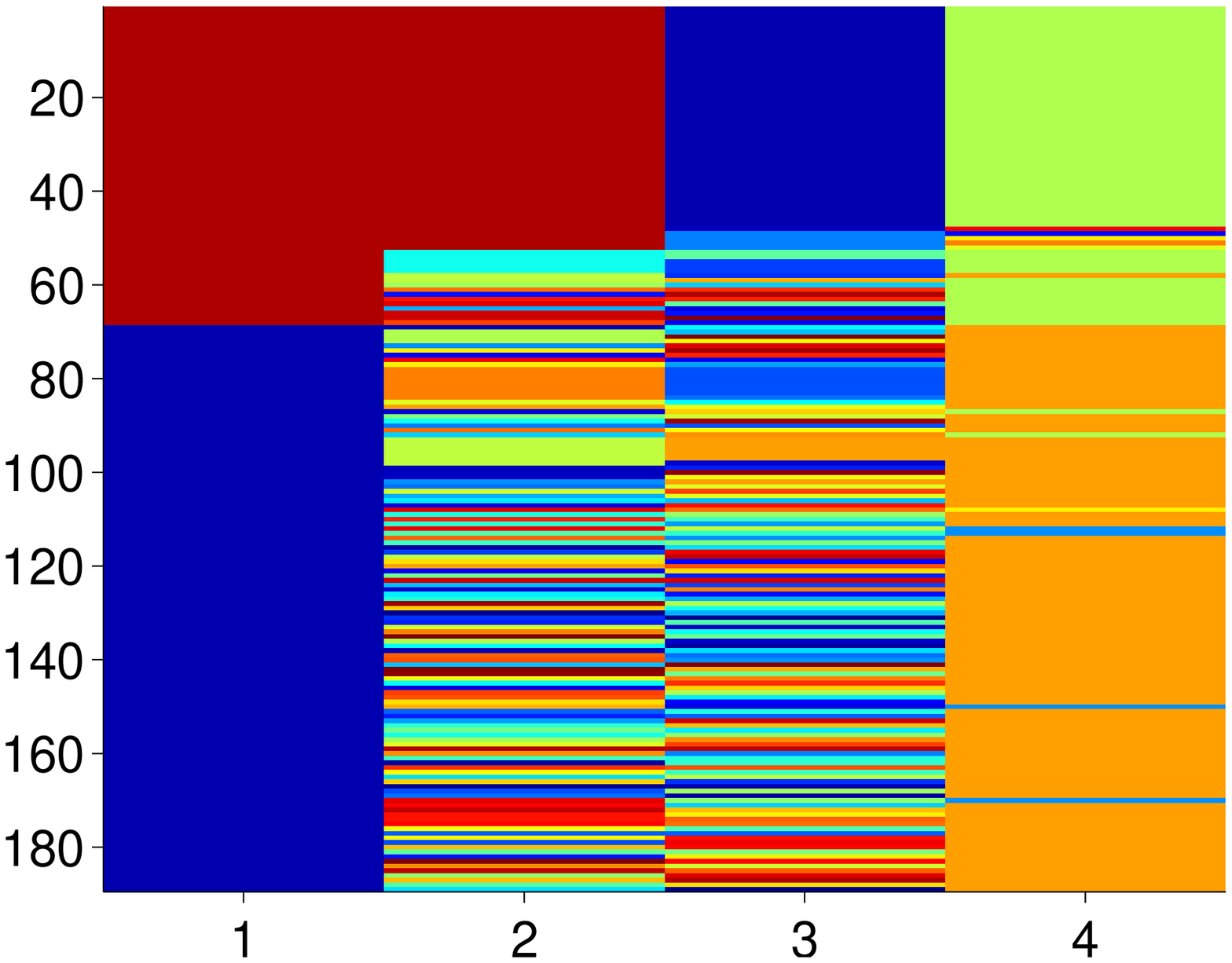}
\caption{(Color) Comparison of the robust groups that we obtained by community detection for each network representation for each of our three case studies of UNGA sessions: (Top) Session 11, (middle) Session 36, and (bottom) Session 58.  For each session, we use a color-coding to visualize the robust groups.  In particular, we compare (1) the partition that we obtained by modularity optimization at default resolution ($\gamma=1$) in the network of voting similarities; (2) the partition that we obtained by modularity optimization of the same network at resolution parameter values of $\gamma \approx 1.360$ (Session 11), $\gamma \approx 1.180$ (Session 36), and $\gamma \approx 1.259$ (Session 58) [each of these values is near the respective right edges of the ranges plotted in Fig.~\ref{plots113658}]; (3) the robust groups that we identified across the indexed points in resolution parameter space in the signed, unipartite representation of voting agreements and disagreements (Fig.~\ref{fig:signed}); and (4) the robust groups of countries that we identified across the indexed points in resolution parameter space in the signed, bipartite network of votes (see Fig.~\ref{fig:bipart}).}


\label{fig:comparison}
\end{figure*}

In Fig.~\ref{fig:comparison}, we compare the community detection results that we obtained partitioning the countries using these different network representations on each of our selected ``case study'' sessions of the UNGA (Sessions 11, 36, and 58).  We consider modularity-optimizing partitions of the network of agreements at two different values of the resolution parameter ($\gamma_1 = 1$ and $\gamma_2 > 1$; see the figure caption for the values of $\gamma_2$ in each session) and illustrate the robust groups that we identified across the selected indexed points in the resolution parameter spaces of the signed, unipartite representation and the signed, bipartite representation (restricting the latter to the identified groups of countries, ignoring the grouping of the resolutions).  For each network representation, we find a dominant voting pattern that includes two large communities corresponding to majority and minority groups.  We observed this for each of the three UNGA sessions in our case studies.  This split appears most clearly for partitions obtained by modularity optimization of the network of agreements at the standard resolution parameter value $\gamma_1 = 1$ (first column) and those obtained by modularity optimization of the bipartite networks (fourth column).  Additionally, one can see (by comparing the two blocks in the first column) small and medium-size cores of groups in the network of agreements that persist for resolution parameter value $\gamma_2 > 1$. For all three sessions, we find such core groups in the second column that arise from each of the large communities in the first column.  In each of the three case-study sessions, a large portion of these core communities remain intact when considering networks of both agreements and disagreements (third column).  These observations appear to be consistent with the expected East-West split of the Cold War and the North-South division of recent sessions that has been described by social scientists using qualitative methods \cite{paulette}.  

In principle, the bipartite network representation would seem to be the best for any investigation of voting, as it contains the complete record of votes cast on resolutions. However, in comparing the identified communities of countries, we observe that the most predominant groupings of the countries also appear prominently when considering the other network representations.  Moreover, in our examples, community detection on the bipartite voting network tends to only return large groups---with few partitions seemingly stable at higher resolution parameter values---whereas the other network representations uncovered a more diverse set of stable groups, including small groups of countries, complementing the information in the large-group partitions. These differences illustrate the importance of considering multiple network representations in the investigation of voting networks and, more generally, that it is crucial to be cognizant of multiple possible network representations when applying such methods.


\section*{Acknowledgements}

We thank Eric Voeten and Adis Merdzanovic for the United Nations voting data \cite{UNData} and James Fowler, Yonatan Lupu, and Mark Newman for useful discussions.  KTM was funded in part by a Summer Undergraduate Research Fellowship from the UNC Office of Undergraduate Research.  PJM \& KTM were additionally funded by the NSF (DMS-0645369).  MAP acknowledges a research award (\#220020177) from the James S. McDonnell Foundation.


\bibliographystyle{apsrev}

\begin{thebibliography}{65}
\expandafter\ifx\csname natexlab\endcsname\relax\def\natexlab#1{#1}\fi
\expandafter\ifx\csname bibnamefont\endcsname\relax
  \def\bibnamefont#1{#1}\fi
\expandafter\ifx\csname bibfnamefont\endcsname\relax
  \def\bibfnamefont#1{#1}\fi
\expandafter\ifx\csname citenamefont\endcsname\relax
  \def\citenamefont#1{#1}\fi
\expandafter\ifx\csname url\endcsname\relax
  \def\url#1{\texttt{#1}}\fi
\expandafter\ifx\csname urlprefix\endcsname\relax\def\urlprefix{URL }\fi
\providecommand{\bibinfo}[2]{#2}
\providecommand{\eprint}[2][]{\url{#2}}

\bibitem[{\citenamefont{Wasserman and Faust}(1994)}]{faust}
\bibinfo{author}{\bibfnamefont{S.}~\bibnamefont{Wasserman}} \bibnamefont{and}
  \bibinfo{author}{\bibfnamefont{K.}~\bibnamefont{Faust}},
  \emph{\bibinfo{title}{Social Network Analysis: Methods and Applications}},
  Structural Analysis in the Social Sciences (\bibinfo{publisher}{Cambridge
  University Press}, \bibinfo{address}{Cambridge, UK}, \bibinfo{year}{1994}).

\bibitem[{\citenamefont{Strogatz}(2001)}]{str01}
\bibinfo{author}{\bibfnamefont{S.~H.} \bibnamefont{Strogatz}},
  \bibinfo{journal}{Nature} \textbf{\bibinfo{volume}{410}},
  \bibinfo{pages}{268} (\bibinfo{year}{2001}).

\bibitem[{\citenamefont{Newman}(2003)}]{Newman2003}
\bibinfo{author}{\bibfnamefont{M.~E.~J.} \bibnamefont{Newman}},
  \bibinfo{journal}{SIAM Review} \textbf{\bibinfo{volume}{45}},
  \bibinfo{pages}{167} (\bibinfo{year}{2003}).

\bibitem[{\citenamefont{Newman}(2010)}]{newman2010}
\bibinfo{author}{\bibfnamefont{M.~E.~J.} \bibnamefont{Newman}},
  \emph{\bibinfo{title}{Networks: An Introduction}} (\bibinfo{publisher}{Oxford
  University Press}, \bibinfo{address}{Oxford, U.K.}, \bibinfo{year}{2010}).

\bibitem[{\citenamefont{Albert and Barab\'{a}si}(2002)}]{albert}
\bibinfo{author}{\bibfnamefont{R.}~\bibnamefont{Albert}} \bibnamefont{and}
  \bibinfo{author}{\bibfnamefont{A.-L.} \bibnamefont{Barab\'{a}si}},
  \bibinfo{journal}{Reviews of Modern Physics} \textbf{\bibinfo{volume}{74}},
  \bibinfo{pages}{47} (\bibinfo{year}{2002}).

\bibitem[{\citenamefont{Dorogovtsev et~al.}(2008)\citenamefont{Dorogovtsev,
  Goltsev, and Mendes}}]{mendes}
\bibinfo{author}{\bibfnamefont{S.~N.} \bibnamefont{Dorogovtsev}},
  \bibinfo{author}{\bibfnamefont{A.~V.} \bibnamefont{Goltsev}},
  \bibnamefont{and} \bibinfo{author}{\bibfnamefont{J.~F.~F.}
  \bibnamefont{Mendes}}, \bibinfo{journal}{Reviews of Modern Physics}
  \textbf{\bibinfo{volume}{80}}, \bibinfo{eid}{1275} (\bibinfo{year}{2008}).

\bibitem[{\citenamefont{Caldarelli}(2007)}]{cald}
\bibinfo{author}{\bibfnamefont{G.}~\bibnamefont{Caldarelli}},
  \emph{\bibinfo{title}{Scale-Free Networks: {C}omplex Webs in Nature and
  Technology}} (\bibinfo{publisher}{Oxford University Press},
  \bibinfo{address}{Oxford, United Kingdom}, \bibinfo{year}{2007}).

\bibitem[{\citenamefont{Fortunato}(2010)}]{santobig}
\bibinfo{author}{\bibfnamefont{S.}~\bibnamefont{Fortunato}},
  \bibinfo{journal}{Physics Reports} \textbf{\bibinfo{volume}{486}},
  \bibinfo{pages}{75 } (\bibinfo{year}{2010}).

\bibitem[{\citenamefont{Porter et~al.}(2009)\citenamefont{Porter, Onnela, and
  Mucha}}]{ourreview}
\bibinfo{author}{\bibfnamefont{M.~A.} \bibnamefont{Porter}},
  \bibinfo{author}{\bibfnamefont{J.-P.} \bibnamefont{Onnela}},
  \bibnamefont{and} \bibinfo{author}{\bibfnamefont{P.~J.} \bibnamefont{Mucha}},
  \bibinfo{journal}{Notices of the American Mathematical Society}
  \textbf{\bibinfo{volume}{56}}, \bibinfo{pages}{1082} (\bibinfo{year}{2009}).

\bibitem[{\citenamefont{Schaeffer}(2007)}]{satu07}
\bibinfo{author}{\bibfnamefont{S.~E.} \bibnamefont{Schaeffer}},
  \bibinfo{journal}{Computer Science Review} \textbf{\bibinfo{volume}{1}},
  \bibinfo{pages}{27} (\bibinfo{year}{2007}).

\bibitem[{\citenamefont{Moody and White}(2003)}]{moody03}
\bibinfo{author}{\bibfnamefont{J.}~\bibnamefont{Moody}} \bibnamefont{and}
  \bibinfo{author}{\bibfnamefont{D.~R.} \bibnamefont{White}},
  \bibinfo{journal}{American Sociological Review}
  \textbf{\bibinfo{volume}{68}}, \bibinfo{pages}{103} (\bibinfo{year}{2003}).

\bibitem[{\citenamefont{Guimer\`{a} and Amaral}(2005)}]{amaral}
\bibinfo{author}{\bibfnamefont{R.}~\bibnamefont{Guimer\`{a}}} \bibnamefont{and}
  \bibinfo{author}{\bibfnamefont{L.~A.~N.} \bibnamefont{Amaral}},
  \bibinfo{journal}{Nature} \textbf{\bibinfo{volume}{433}},
  \bibinfo{pages}{895} (\bibinfo{year}{2005}).

\bibitem[{\citenamefont{Lewis et~al.}(2010)\citenamefont{Lewis, Jones, Porter,
  and Deane}}]{anna}
\bibinfo{author}{\bibfnamefont{A.~C.~F.} \bibnamefont{Lewis}},
  \bibinfo{author}{\bibfnamefont{N.~S.} \bibnamefont{Jones}},
  \bibinfo{author}{\bibfnamefont{M.~A.} \bibnamefont{Porter}},
  \bibnamefont{and} \bibinfo{author}{\bibfnamefont{C.~M.} \bibnamefont{Deane}},
  \bibinfo{journal}{BMC Systems Biology} \textbf{\bibinfo{volume}{4}}
  (\bibinfo{year}{2010}).

\bibitem[{\citenamefont{Girvan and Newman}(2002)}]{structpnas}
\bibinfo{author}{\bibfnamefont{M.}~\bibnamefont{Girvan}} \bibnamefont{and}
  \bibinfo{author}{\bibfnamefont{M.~E.~J.} \bibnamefont{Newman}},
  \bibinfo{journal}{Proceedings of the National Academy of Sciences}
  \textbf{\bibinfo{volume}{99}}, \bibinfo{pages}{7821} (\bibinfo{year}{2002}).

\bibitem[{\citenamefont{Callaghan et~al.}(2007)\citenamefont{Callaghan, Mucha,
  and Porter}}]{bcs}
\bibinfo{author}{\bibfnamefont{T.}~\bibnamefont{Callaghan}},
  \bibinfo{author}{\bibfnamefont{P.~J.} \bibnamefont{Mucha}}, \bibnamefont{and}
  \bibinfo{author}{\bibfnamefont{M.~A.} \bibnamefont{Porter}},
  \bibinfo{journal}{American Mathematical Monthly}
  \textbf{\bibinfo{volume}{114}}, \bibinfo{pages}{761} (\bibinfo{year}{2007}).

\bibitem[{\citenamefont{Porter et~al.}(2005)\citenamefont{Porter, Mucha,
  Newman, and Warmbrand}}]{congshort}
\bibinfo{author}{\bibfnamefont{M.~A.} \bibnamefont{Porter}},
  \bibinfo{author}{\bibfnamefont{P.~J.} \bibnamefont{Mucha}},
  \bibinfo{author}{\bibfnamefont{M.~E.~J.} \bibnamefont{Newman}},
  \bibnamefont{and} \bibinfo{author}{\bibfnamefont{C.~M.}
  \bibnamefont{Warmbrand}}, \bibinfo{journal}{Proceedings of the National
  Academy of Sciences} \textbf{\bibinfo{volume}{102}}, \bibinfo{pages}{7057}
  (\bibinfo{year}{2005}).

\bibitem[{\citenamefont{Porter et~al.}(2007)\citenamefont{Porter, Mucha,
  Newman, and Friend}}]{conglong}
\bibinfo{author}{\bibfnamefont{M.~A.} \bibnamefont{Porter}},
  \bibinfo{author}{\bibfnamefont{P.~J.} \bibnamefont{Mucha}},
  \bibinfo{author}{\bibfnamefont{M.~E.~J.} \bibnamefont{Newman}},
  \bibnamefont{and} \bibinfo{author}{\bibfnamefont{A.~J.}
  \bibnamefont{Friend}}, \bibinfo{journal}{Physica A}
  \textbf{\bibinfo{volume}{386}}, \bibinfo{pages}{414} (\bibinfo{year}{2007}).

\bibitem[{\citenamefont{Porter et~al.}(2006)\citenamefont{Porter, Friend,
  Mucha, and Newman}}]{conggallery}
\bibinfo{author}{\bibfnamefont{M.~A.} \bibnamefont{Porter}},
  \bibinfo{author}{\bibfnamefont{A.~J.} \bibnamefont{Friend}},
  \bibinfo{author}{\bibfnamefont{P.~J.} \bibnamefont{Mucha}}, \bibnamefont{and}
  \bibinfo{author}{\bibfnamefont{M.~E.~J.} \bibnamefont{Newman}},
  \bibinfo{journal}{Chaos} \textbf{\bibinfo{volume}{16}},
  \bibinfo{pages}{041106} (\bibinfo{year}{2006}).

\bibitem[{\citenamefont{Zhang et~al.}(2008)\citenamefont{Zhang, Friend, Traud,
  Porter, Fowler, and Mucha}}]{yan}
\bibinfo{author}{\bibfnamefont{Y.}~\bibnamefont{Zhang}},
  \bibinfo{author}{\bibfnamefont{A.~J.} \bibnamefont{Friend}},
  \bibinfo{author}{\bibfnamefont{A.~L.} \bibnamefont{Traud}},
  \bibinfo{author}{\bibfnamefont{M.~A.} \bibnamefont{Porter}},
  \bibinfo{author}{\bibfnamefont{J.~H.} \bibnamefont{Fowler}},
  \bibnamefont{and} \bibinfo{author}{\bibfnamefont{P.~J.} \bibnamefont{Mucha}},
  \bibinfo{journal}{Physica A} \textbf{\bibinfo{volume}{387}},
  \bibinfo{pages}{1705} (\bibinfo{year}{2008}).

\bibitem[{\citenamefont{Waugh et~al.}(2010)\citenamefont{Waugh, Pei, Fowler,
  Mucha, and Porter}}]{waugh}
\bibinfo{author}{\bibfnamefont{A.~S.} \bibnamefont{Waugh}},
  \bibinfo{author}{\bibfnamefont{L.}~\bibnamefont{Pei}},
  \bibinfo{author}{\bibfnamefont{J.~H.} \bibnamefont{Fowler}},
  \bibinfo{author}{\bibfnamefont{P.~J.} \bibnamefont{Mucha}}, \bibnamefont{and}
  \bibinfo{author}{\bibfnamefont{M.~A.} \bibnamefont{Porter}}
  (\bibinfo{year}{2010}), \bibinfo{note}{arXiv:0907.3509}.

\bibitem[{\citenamefont{Gonz\'{a}lez et~al.}(2007)\citenamefont{Gonz\'{a}lez,
  Herrmann, Kert\'{e}sz, and Vicsek}}]{marta}
\bibinfo{author}{\bibfnamefont{M.~C.} \bibnamefont{Gonz\'{a}lez}},
  \bibinfo{author}{\bibfnamefont{H.~J.} \bibnamefont{Herrmann}},
  \bibinfo{author}{\bibfnamefont{J.}~\bibnamefont{Kert\'{e}sz}},
  \bibnamefont{and} \bibinfo{author}{\bibfnamefont{T.}~\bibnamefont{Vicsek}},
  \bibinfo{journal}{Physica A} \textbf{\bibinfo{volume}{379}},
  \bibinfo{pages}{307} (\bibinfo{year}{2007}).

\bibitem[{\citenamefont{Traud et~al.}(2010)\citenamefont{Traud, Kelsic, Mucha,
  and Porter}}]{facebook}
\bibinfo{author}{\bibfnamefont{A.~L.} \bibnamefont{Traud}},
  \bibinfo{author}{\bibfnamefont{E.~D.} \bibnamefont{Kelsic}},
  \bibinfo{author}{\bibfnamefont{P.~J.} \bibnamefont{Mucha}}, \bibnamefont{and}
  \bibinfo{author}{\bibfnamefont{M.~A.} \bibnamefont{Porter}}
  (\bibinfo{year}{2010}), \bibinfo{note}{{SIAM} Review, in press
  (arXiv:0809.0960)}.

\bibitem[{\citenamefont{Onnela et~al.}(2007)\citenamefont{Onnela, Saram\"{a}ki,
  Hyv\"{o}nen, Szab\'{o}, Lazer, Kaski, Kert\'{e}sz, and Barab\'{a}si}}]{jp}
\bibinfo{author}{\bibfnamefont{J.-P.} \bibnamefont{Onnela}},
  \bibinfo{author}{\bibfnamefont{J.}~\bibnamefont{Saram\"{a}ki}},
  \bibinfo{author}{\bibfnamefont{J.}~\bibnamefont{Hyv\"{o}nen}},
  \bibinfo{author}{\bibfnamefont{G.}~\bibnamefont{Szab\'{o}}},
  \bibinfo{author}{\bibfnamefont{D.}~\bibnamefont{Lazer}},
  \bibinfo{author}{\bibfnamefont{K.}~\bibnamefont{Kaski}},
  \bibinfo{author}{\bibfnamefont{J.}~\bibnamefont{Kert\'{e}sz}},
  \bibnamefont{and} \bibinfo{author}{\bibfnamefont{A.-L.}
  \bibnamefont{Barab\'{a}si}}, \bibinfo{journal}{Proceedings of the National
  Academy of Sciences} \textbf{\bibinfo{volume}{104}}, \bibinfo{pages}{7332}
  (\bibinfo{year}{2007}).

\bibitem[{\citenamefont{Palla et~al.}(2007)\citenamefont{Palla, Barab\'{a}si,
  and Vicsek}}]{palla07}
\bibinfo{author}{\bibfnamefont{G.}~\bibnamefont{Palla}},
  \bibinfo{author}{\bibfnamefont{A.-L.} \bibnamefont{Barab\'{a}si}},
  \bibnamefont{and} \bibinfo{author}{\bibfnamefont{T.}~\bibnamefont{Vicsek}},
  \bibinfo{journal}{Nature} \textbf{\bibinfo{volume}{446}},
  \bibinfo{pages}{664} (\bibinfo{year}{2007}).

\bibitem[{\citenamefont{Fenn et~al.}(2009)\citenamefont{Fenn, Porter, McDonald,
  Williams, Johnson, and Jones}}]{fenn}
\bibinfo{author}{\bibfnamefont{D.~J.} \bibnamefont{Fenn}},
  \bibinfo{author}{\bibfnamefont{M.~A.} \bibnamefont{Porter}},
  \bibinfo{author}{\bibfnamefont{M.}~\bibnamefont{McDonald}},
  \bibinfo{author}{\bibfnamefont{S.}~\bibnamefont{Williams}},
  \bibinfo{author}{\bibfnamefont{N.~F.} \bibnamefont{Johnson}},
  \bibnamefont{and} \bibinfo{author}{\bibfnamefont{N.~S.} \bibnamefont{Jones}},
  \bibinfo{journal}{Chaos} \textbf{\bibinfo{volume}{19}},
  \bibinfo{pages}{033119} (\bibinfo{year}{2009}).

\bibitem[{\citenamefont{Fenn et~al.}(2010)\citenamefont{Fenn, Porter, Mucha,
  McDonald, Williams, Johnson, and Jones}}]{fennlong}
\bibinfo{author}{\bibfnamefont{D.~J.} \bibnamefont{Fenn}},
  \bibinfo{author}{\bibfnamefont{M.~A.} \bibnamefont{Porter}},
  \bibinfo{author}{\bibfnamefont{P.~J.} \bibnamefont{Mucha}},
  \bibinfo{author}{\bibfnamefont{M.}~\bibnamefont{McDonald}},
  \bibinfo{author}{\bibfnamefont{S.}~\bibnamefont{Williams}},
  \bibinfo{author}{\bibfnamefont{N.~F.} \bibnamefont{Johnson}},
  \bibnamefont{and} \bibinfo{author}{\bibfnamefont{N.~S.} \bibnamefont{Jones}}
  (\bibinfo{year}{2010}), \bibinfo{note}{arXiv:0905.4912}.

\bibitem[{\citenamefont{Mucha et~al.}(2010)\citenamefont{Mucha, Richardson,
  Macon, Porter, and Onnela}}]{multislice}
\bibinfo{author}{\bibfnamefont{P.~J.} \bibnamefont{Mucha}},
  \bibinfo{author}{\bibfnamefont{T.}~\bibnamefont{Richardson}},
  \bibinfo{author}{\bibfnamefont{K.}~\bibnamefont{Macon}},
  \bibinfo{author}{\bibfnamefont{M.~A.} \bibnamefont{Porter}},
  \bibnamefont{and} \bibinfo{author}{\bibfnamefont{J.-P.}
  \bibnamefont{Onnela}}, \bibinfo{journal}{Science}
  \textbf{\bibinfo{volume}{328}}, \bibinfo{pages}{876} (\bibinfo{year}{2010}).

\bibitem[{\citenamefont{G\'{o}mez et~al.}(2009)\citenamefont{G\'{o}mez, Jensen,
  and Arenas}}]{gomez09}
\bibinfo{author}{\bibfnamefont{S.}~\bibnamefont{G\'{o}mez}},
  \bibinfo{author}{\bibfnamefont{P.}~\bibnamefont{Jensen}}, \bibnamefont{and}
  \bibinfo{author}{\bibfnamefont{A.}~\bibnamefont{Arenas}},
  \bibinfo{journal}{Physical Review E} \textbf{\bibinfo{volume}{80}},
  \bibinfo{eid}{016114} (pages~\bibinfo{numpages}{5}) (\bibinfo{year}{2009}).

\bibitem[{\citenamefont{Traag and Bruggeman}(2009)}]{traag2009}
\bibinfo{author}{\bibfnamefont{V.~A.} \bibnamefont{Traag}} \bibnamefont{and}
  \bibinfo{author}{\bibfnamefont{J.}~\bibnamefont{Bruggeman}},
  \bibinfo{journal}{Physical Review E} \textbf{\bibinfo{volume}{80}},
  \bibinfo{pages}{036115} (\bibinfo{year}{2009}).

\bibitem[{\citenamefont{Lijphart}(1963)}]{arend}
\bibinfo{author}{\bibfnamefont{A.}~\bibnamefont{Lijphart}},
  \bibinfo{journal}{American Political Science Review}
  \textbf{\bibinfo{volume}{57}}, \bibinfo{pages}{902} (\bibinfo{year}{1963}).

\bibitem[{\citenamefont{Hafner-Burton et~al.}(2009)\citenamefont{Hafner-Burton,
  Kahler, and Montgomery}}]{burton09}
\bibinfo{author}{\bibfnamefont{E.~M.} \bibnamefont{Hafner-Burton}},
  \bibinfo{author}{\bibfnamefont{M.}~\bibnamefont{Kahler}}, \bibnamefont{and}
  \bibinfo{author}{\bibfnamefont{A.~H.} \bibnamefont{Montgomery}},
  \bibinfo{journal}{International Organization} \textbf{\bibinfo{volume}{63}},
  \bibinfo{pages}{559} (\bibinfo{year}{2009}).

\bibitem[{\citenamefont{Pevehouse et~al.}(2004)\citenamefont{Pevehouse,
  Nordstrom, and Warnke}}]{cow2}
\bibinfo{author}{\bibfnamefont{J.}~\bibnamefont{Pevehouse}},
  \bibinfo{author}{\bibfnamefont{T.}~\bibnamefont{Nordstrom}},
  \bibnamefont{and} \bibinfo{author}{\bibfnamefont{K.}~\bibnamefont{Warnke}},
  \bibinfo{journal}{Conflict Management and Peace Science}
  \textbf{\bibinfo{volume}{21}}, \bibinfo{pages}{101} (\bibinfo{year}{2004}).

\bibitem[{\citenamefont{Diehl}(2009)}]{cow}
\bibinfo{author}{\bibfnamefont{P.}~\bibnamefont{Diehl}} (\bibinfo{year}{2009}),
  \bibinfo{note}{available at \url{http://www.correlatesofwar.org/}}.

\bibitem[{\citenamefont{Cranmer and Siverson}(2008)}]{cranmer08}
\bibinfo{author}{\bibfnamefont{S.~J.} \bibnamefont{Cranmer}} \bibnamefont{and}
  \bibinfo{author}{\bibfnamefont{R.~M.} \bibnamefont{Siverson}},
  \bibinfo{journal}{Journal of Politics} \textbf{\bibinfo{volume}{70}},
  \bibinfo{pages}{794} (\bibinfo{year}{2008}).

\bibitem[{\citenamefont{Voeten}(2000)}]{voeten}
\bibinfo{author}{\bibfnamefont{E.}~\bibnamefont{Voeten}},
  \bibinfo{journal}{International Organization} pp. \bibinfo{pages}{185--215}
  (\bibinfo{year}{2000}).

\bibitem[{\citenamefont{Poole and Rosenthal}(1997)}]{pr97}
\bibinfo{author}{\bibfnamefont{K.~T.} \bibnamefont{Poole}} \bibnamefont{and}
  \bibinfo{author}{\bibfnamefont{H.}~\bibnamefont{Rosenthal}},
  \emph{\bibinfo{title}{Congress: A Political-Economic History of Roll Call
  Voting}} (\bibinfo{publisher}{Oxford University Press},
  \bibinfo{address}{Oxford, United Kingdom}, \bibinfo{year}{1997}).

\bibitem[{\citenamefont{Riss}(2010)}]{voteworld}
\bibinfo{author}{\bibfnamefont{H.}~\bibnamefont{Riss}} (\bibinfo{year}{2010}),
  \bibinfo{note}{\url{http://voteworld.berkeley.edu/}}.

\bibitem[{\citenamefont{Lloyd}(2008)}]{paulette}
\bibinfo{author}{\bibfnamefont{P.}~\bibnamefont{Lloyd}} (\bibinfo{year}{2008}),
  \bibinfo{note}{unpublished}.

\bibitem[{\citenamefont{{United Nations}}(2009)}]{UNGA}
\bibinfo{author}{\bibnamefont{{United Nations}}} (\bibinfo{year}{2009}),
  \bibinfo{note}{available at \url{http://www.un.org/ga/}}.

\bibitem[{\citenamefont{Russet}(1966)}]{russett}
\bibinfo{author}{\bibfnamefont{B.~M.} \bibnamefont{Russet}},
  \bibinfo{journal}{American Political Science Review}
  \textbf{\bibinfo{volume}{60}}, \bibinfo{pages}{327} (\bibinfo{year}{1966}).

\bibitem[{\citenamefont{Fowler}(2006{\natexlab{a}})}]{fowlershort}
\bibinfo{author}{\bibfnamefont{J.~H.} \bibnamefont{Fowler}},
  \bibinfo{journal}{Social Networks} \textbf{\bibinfo{volume}{28}},
  \bibinfo{pages}{456} (\bibinfo{year}{2006}{\natexlab{a}}).

\bibitem[{\citenamefont{Fowler}(2006{\natexlab{b}})}]{fowler}
\bibinfo{author}{\bibfnamefont{J.~H.} \bibnamefont{Fowler}},
  \bibinfo{journal}{Political Analysis} \textbf{\bibinfo{volume}{14}},
  \bibinfo{pages}{454} (\bibinfo{year}{2006}{\natexlab{b}}).

\bibitem[{\citenamefont{Gartzke}(2009)}]{gartzke}
\bibinfo{author}{\bibfnamefont{E.}~\bibnamefont{Gartzke}}
  (\bibinfo{year}{2009}), \bibinfo{note}{available at
  \url{http://dss.ucsd.edu/~egartzke}}.

\bibitem[{\citenamefont{Signorino and Ritter}(1999)}]{signorino}
\bibinfo{author}{\bibfnamefont{S.}~\bibnamefont{Signorino}} \bibnamefont{and}
  \bibinfo{author}{\bibfnamefont{J.}~\bibnamefont{Ritter}},
  \bibinfo{journal}{International Studies Quarterly}
  \textbf{\bibinfo{volume}{43}}, \bibinfo{pages}{115} (\bibinfo{year}{1999}).

\bibitem[{\citenamefont{Sweeney and Keshk}(2009)}]{sweeney}
\bibinfo{author}{\bibfnamefont{K.}~\bibnamefont{Sweeney}} \bibnamefont{and}
  \bibinfo{author}{\bibfnamefont{O.~M.~G.} \bibnamefont{Keshk}},
  \bibinfo{journal}{Conflict Management and Peace Science}
  \textbf{\bibinfo{volume}{22}}, \bibinfo{pages}{165} (\bibinfo{year}{2009}).

\bibitem[{\citenamefont{Cohen}(1960)}]{cohen}
\bibinfo{author}{\bibfnamefont{J.}~\bibnamefont{Cohen}},
  \bibinfo{journal}{Educational and Psychological Measurement}
  \textbf{\bibinfo{volume}{20}}, \bibinfo{pages}{37} (\bibinfo{year}{1960}).

\bibitem[{\citenamefont{Heimo et~al.}(2008)\citenamefont{Heimo, Kumpula, Kaski,
  and Saram\"{a}ki}}]{heimo}
\bibinfo{author}{\bibfnamefont{T.}~\bibnamefont{Heimo}},
  \bibinfo{author}{\bibfnamefont{J.~S.} \bibnamefont{Kumpula}},
  \bibinfo{author}{\bibfnamefont{K.}~\bibnamefont{Kaski}}, \bibnamefont{and}
  \bibinfo{author}{\bibfnamefont{J.}~\bibnamefont{Saram\"{a}ki}},
  \bibinfo{journal}{Journal of Statistical Mechics}
  \textbf{\bibinfo{volume}{2008}}, \bibinfo{pages}{P08007}
  (\bibinfo{year}{2008}).

\bibitem[{\citenamefont{Newman and Girvan}(2003)}]{structmix}
\bibinfo{author}{\bibfnamefont{M.~E.~J.} \bibnamefont{Newman}}
  \bibnamefont{and} \bibinfo{author}{\bibfnamefont{M.}~\bibnamefont{Girvan}},
  in \emph{\bibinfo{booktitle}{Statistical Mechanics of Complex Networks}},
  edited by \bibinfo{editor}{\bibfnamefont{R.}~\bibnamefont{Pastor-Satorras}},
  \bibinfo{editor}{\bibfnamefont{J.}~\bibnamefont{Rubi}}, \bibnamefont{and}
  \bibinfo{editor}{\bibfnamefont{A.}~\bibnamefont{Diaz-Guilera}}
  (\bibinfo{publisher}{Springer-Verlag}, \bibinfo{address}{Berlin, Germany},
  \bibinfo{year}{2003}).

\bibitem[{\citenamefont{Newman and Girvan}(2004)}]{structeval}
\bibinfo{author}{\bibfnamefont{M.~E.~J.} \bibnamefont{Newman}}
  \bibnamefont{and} \bibinfo{author}{\bibfnamefont{M.}~\bibnamefont{Girvan}},
  \bibinfo{journal}{Physical Review E} \textbf{\bibinfo{volume}{69}}
  (\bibinfo{year}{2004}).

\bibitem[{\citenamefont{Newman}(2006{\natexlab{a}})}]{newmodlong}
\bibinfo{author}{\bibfnamefont{M.~E.~J.} \bibnamefont{Newman}},
  \bibinfo{journal}{Physical Review E} \textbf{\bibinfo{volume}{74}},
  \bibinfo{pages}{036104} (\bibinfo{year}{2006}{\natexlab{a}}).

\bibitem[{\citenamefont{Newman}(2006{\natexlab{b}})}]{newmod}
\bibinfo{author}{\bibfnamefont{M.~E.~J.} \bibnamefont{Newman}},
  \bibinfo{journal}{Proceedings of the National Academy of Sciences}
  \textbf{\bibinfo{volume}{103}}, \bibinfo{pages}{8577}
  (\bibinfo{year}{2006}{\natexlab{b}}).

\bibitem[{\citenamefont{Richardson et~al.}(2009)\citenamefont{Richardson,
  Mucha, and Porter}}]{richardson}
\bibinfo{author}{\bibfnamefont{T.}~\bibnamefont{Richardson}},
  \bibinfo{author}{\bibfnamefont{P.~J.} \bibnamefont{Mucha}}, \bibnamefont{and}
  \bibinfo{author}{\bibfnamefont{M.~A.} \bibnamefont{Porter}},
  \bibinfo{journal}{Physical Review E} \textbf{\bibinfo{volume}{80}},
  \bibinfo{pages}{036111} (\bibinfo{year}{2009}).

\bibitem[{\citenamefont{Blondel et~al.}(2008)\citenamefont{Blondel, Guillaume,
  Lambiotte, and Lefebvre}}]{blondel}
\bibinfo{author}{\bibfnamefont{V.~D.} \bibnamefont{Blondel}},
  \bibinfo{author}{\bibfnamefont{J.-L.} \bibnamefont{Guillaume}},
  \bibinfo{author}{\bibfnamefont{R.}~\bibnamefont{Lambiotte}},
  \bibnamefont{and} \bibinfo{author}{\bibfnamefont{E.}~\bibnamefont{Lefebvre}},
  \bibinfo{journal}{Journal of Statistical Mechanics} p.
  \bibinfo{pages}{P10008} (\bibinfo{year}{2008}).

\bibitem[{\citenamefont{Kernighan and Lin}(1970)}]{kl}
\bibinfo{author}{\bibfnamefont{B.~W.} \bibnamefont{Kernighan}}
  \bibnamefont{and} \bibinfo{author}{\bibfnamefont{S.}~\bibnamefont{Lin}},
  \bibinfo{journal}{The Bell System Technical Journal}
  \textbf{\bibinfo{volume}{49}}, \bibinfo{pages}{291} (\bibinfo{year}{1970}).

\bibitem[{\citenamefont{Danon et~al.}(2005)\citenamefont{Danon, Diaz-Guilera,
  Duch, and Arenas}}]{commreview}
\bibinfo{author}{\bibfnamefont{L.}~\bibnamefont{Danon}},
  \bibinfo{author}{\bibfnamefont{A.}~\bibnamefont{Diaz-Guilera}},
  \bibinfo{author}{\bibfnamefont{J.}~\bibnamefont{Duch}}, \bibnamefont{and}
  \bibinfo{author}{\bibfnamefont{A.}~\bibnamefont{Arenas}},
  \bibinfo{journal}{Journal of Statistical Mechanics}  (\bibinfo{year}{2005}).

\bibitem[{\citenamefont{Good et~al.}(2010)\citenamefont{Good, de~Montjoye, and
  Clauset}}]{good2010}
\bibinfo{author}{\bibfnamefont{B.~H.} \bibnamefont{Good}},
  \bibinfo{author}{\bibfnamefont{Y.-A.} \bibnamefont{de~Montjoye}},
  \bibnamefont{and} \bibinfo{author}{\bibfnamefont{A.}~\bibnamefont{Clauset}},
  \bibinfo{journal}{Physical Review E} \textbf{\bibinfo{volume}{81}},
  \bibinfo{pages}{046106} (\bibinfo{year}{2010}).

\bibitem[{\citenamefont{Fortunato and Barth\'{e}lemy}(2007)}]{resolution}
\bibinfo{author}{\bibfnamefont{S.}~\bibnamefont{Fortunato}} \bibnamefont{and}
  \bibinfo{author}{\bibfnamefont{M.}~\bibnamefont{Barth\'{e}lemy}},
  \bibinfo{journal}{Proceedings of the National Academy of Sciences}
  \textbf{\bibinfo{volume}{104}}, \bibinfo{pages}{36} (\bibinfo{year}{2007}).

\bibitem[{\citenamefont{Reichardt and Bornholdt}(2006)}]{spinglass}
\bibinfo{author}{\bibfnamefont{J.}~\bibnamefont{Reichardt}} \bibnamefont{and}
  \bibinfo{author}{\bibfnamefont{S.}~\bibnamefont{Bornholdt}},
  \bibinfo{journal}{Physical Review E} \textbf{\bibinfo{volume}{74}},
  \bibinfo{pages}{016110} (\bibinfo{year}{2006}).

\bibitem[{\citenamefont{Brandes et~al.}(2008)\citenamefont{Brandes, Delling,
  Gaertler, Goerke, Hoefer, Nikoloski, and Wagner}}]{np}
\bibinfo{author}{\bibfnamefont{U.}~\bibnamefont{Brandes}},
  \bibinfo{author}{\bibfnamefont{D.}~\bibnamefont{Delling}},
  \bibinfo{author}{\bibfnamefont{M.}~\bibnamefont{Gaertler}},
  \bibinfo{author}{\bibfnamefont{R.}~\bibnamefont{Goerke}},
  \bibinfo{author}{\bibfnamefont{M.}~\bibnamefont{Hoefer}},
  \bibinfo{author}{\bibfnamefont{Z.}~\bibnamefont{Nikoloski}},
  \bibnamefont{and} \bibinfo{author}{\bibfnamefont{D.}~\bibnamefont{Wagner}},
  \bibinfo{journal}{IEEE Transactions on Knowledge and Data Engineering}
  \textbf{\bibinfo{volume}{20}}, \bibinfo{pages}{172} (\bibinfo{year}{2008}).

\bibitem[{\citenamefont{Bonacich and Lloyd}(2004)}]{status}
\bibinfo{author}{\bibfnamefont{P.}~\bibnamefont{Bonacich}} \bibnamefont{and}
  \bibinfo{author}{\bibfnamefont{P.}~\bibnamefont{Lloyd}},
  \bibinfo{journal}{Social Networks} \textbf{\bibinfo{volume}{26}},
  \bibinfo{pages}{331} (\bibinfo{year}{2004}).

\bibitem[{\citenamefont{Szell et~al.}(2010)\citenamefont{Szell, Lambiotte, and
  Thurner}}]{renaud10}
\bibinfo{author}{\bibfnamefont{M.}~\bibnamefont{Szell}},
  \bibinfo{author}{\bibfnamefont{R.}~\bibnamefont{Lambiotte}},
  \bibnamefont{and} \bibinfo{author}{\bibfnamefont{S.}~\bibnamefont{Thurner}},
  \bibinfo{journal}{Proceedings of the National Academy of Sciences}
  \textbf{\bibinfo{volume}{107}}, \bibinfo{pages}{13636}
  (\bibinfo{year}{2010}).

\bibitem[{\citenamefont{Barber}(2007)}]{barber2007}
\bibinfo{author}{\bibfnamefont{M.~J.} \bibnamefont{Barber}},
  \bibinfo{journal}{Physical Review E} \textbf{\bibinfo{volume}{76}},
  \bibinfo{eid}{066102} (\bibinfo{year}{2007}).

\bibitem[{\citenamefont{{The Group of 77}}(2009)}]{g77}
\bibinfo{author}{\bibnamefont{{The Group of 77}}} (\bibinfo{year}{2009}),
  \bibinfo{note}{available at \url{http://www.g77.org/doc/members.html}}.

\bibitem[{\citenamefont{Onnela et~al.}(2010)\citenamefont{Onnela, Fenn, Reid,
  Porter, Mucha, Fricker, and Jones}}]{reid}
\bibinfo{author}{\bibfnamefont{J.-P.} \bibnamefont{Onnela}},
  \bibinfo{author}{\bibfnamefont{D.~J.} \bibnamefont{Fenn}},
  \bibinfo{author}{\bibfnamefont{S.}~\bibnamefont{Reid}},
  \bibinfo{author}{\bibfnamefont{M.~A.} \bibnamefont{Porter}},
  \bibinfo{author}{\bibfnamefont{P.~J.} \bibnamefont{Mucha}},
  \bibinfo{author}{\bibfnamefont{M.~D.} \bibnamefont{Fricker}},
  \bibnamefont{and} \bibinfo{author}{\bibfnamefont{N.~S.} \bibnamefont{Jones}}
  (\bibinfo{year}{2010}), \bibinfo{note}{arXiv:1006.5731}.

\bibitem[{\citenamefont{Voeten and Merdzanovic}(2009)}]{UNData}
\bibinfo{author}{\bibfnamefont{E.}~\bibnamefont{Voeten}} \bibnamefont{and}
  \bibinfo{author}{\bibfnamefont{A.}~\bibnamefont{Merdzanovic}}
  (\bibinfo{year}{2009}), \bibinfo{note}{available at
  \url{http://hdl.handle.net/1902.1/12379}}.

\end{thebibliography}


\clearpage{}

\begin{table*}[htp]
\caption{39 Countries partitioned in the ``North'' community consistently in Sessions 46--63 (cf.\ Fig.~\ref{g77}). 
}\label{NorthBloc}
\begin{tabular}{l l l l l}
\hline
Albania	&	Estonia	&	Israel		&	Marshall Islands	&	Spain	\\
Armenia	&	Finland	&	Italy			&	Netherlands		&	Sweden	\\
Australia	&	France	&	Japan		&	New Zealand		&	Turkey	\\
Austria	&	Germany	&	Latvia		&	Norway			&	Ukraine	\\
Belgium	&	Greece	&	Liechtenstein	&	Poland			&	United Kingdom	\\
Bulgaria	&	Hungary	&	Lithuania		&	Portugal			&	United States of America	\\
Canada	&	Iceland	&	Luxembourg	&	Romania			&	Western Samoa	\\
Denmark	&	Ireland	&	Malta		&	South Korea		&		\\
\end{tabular}
\end{table*}

\begin{table*}[htp]
\caption{84 Countries partitioned in the ``South'' community consistently in Sessions 46--63 (cf.\ Fig.~\ref{g77}).}\label{SouthBloc}
\begin{tabular}{l l l l l l}
\hline
Afghanistan	&Chile		&Guinea		&Maldives	&Qatar		&Trinidad and Tobago  \\
Algeria		&China		&Guyana		&Mali		&Rwanda		&Tunisia  \\
Angola		&Colombia	&India		&Mauritius	&Saudi Arabia	&Uganda	\\
Bahrain		&Comoros	&Indonesia	&Mexico		&Senegal		&United Arab Emirates  \\
Bangladesh	&Congo		&Iran		&Morocco		&Sierra Leone	&Venezuela  \\
Belize		&Costa Rica	&Jamaica		&Mozambique	&Singapore	&Vietnam	\\
Benin		&Cote D'Ivoire	&Jordan		&Myanmar	&Sri Lanka	&Yemen	\\
Bhutan		&Cuba		&Kenya		&Namibia		&St. Lucia		&Zambia	\\
Bolivia		&Djibouti		&Kuwait		&Nepal		&Sudan		&Zimbabwe  \\
Botswana		&Ecuador		&Laos		&Nigeria		&Surinam		&		\\
Brazil		&Egypt		&Lebanon	&North Korea	&Swaziland	&		\\
Brunei		&El Salvador	&Lesotho		&Oman		&Syria		&		\\
Burkina Faso	&Ethiopia		&Libya		&Pakistan		&Tanzania	&		\\
Cameroon	&Gabon		&Madagascar	&Peru		&Thailand		&		\\
Cape Verde	&Ghana		&Malaysia	&Philippines	&Togo		&		\\
\end{tabular}
\end{table*}

\begin{table*}[htp]
\begin{center}
\caption{Countries in the two large robust groups of the Session 11 voting agreement network (which we highlight in Fig.~\ref{plots113658} using dashed lines).  
}\label{subgroup}
\begin{tabular}{l || l | l l }
Western Core Community & \multicolumn{3}{c}{Eastern Community} \\
& ``No"-Voting Community & \multicolumn{2}{c}{Abstaining Community} \\
\hline
United Kingdom 	& Poland 			&Yugoslavia 	& Morocco \\
Netherlands 		& Hungary 		& Libya 		& Sudan\\
Belgium 			& Czechoslovakia 	& Egypt 		& Syria\\
Luxembourg 		& Albania 			& Lebanon 	& Jordan\\
France 			& Bulgaria 		& Saudi Arabia & Yemen \\
Portugal 			& Romania 		& Afghanistan 	& India \\
South Africa 		& Russia 			& Myanmar 	& Sri Lanka \\
Israel 			& Ukraine 		& Indonesia 			\\
Australia  			& Belarus			& \\
New Zealand 		& 				& \\
\end{tabular}
\end{center}
\end{table*}

\begin{table*}[htp]
\caption{Countries in the two large robust groups of the Session 36 voting agreement network (which we highlight in Fig.~\ref{plots113658} using dashed lines).
}
\label{36table}
\begin{tabular}{l l l | l l}
\multicolumn{3}{c}{Western Community} \vline & \multicolumn{2}{c}{Eastern Community} \\
\hline
USA 		& Canada 		& UK 	& Cuba 			& East Germany \\
Guatemala 	& Paraguay 	& Belize 	& Poland 			& Hungary \\
Ireland 		& Netherlands 		&Belgium & Czechoslovakia 	& Albania \\
Luxembourg & France 			& Spain & Bulgaria 			& Russia \\
Portugal 	& West Germany 	& Austria & Ukraine 			& Belarus	\\
Italy 		& Greece 			& Turkey & Seychelles 		& \\
Sweden		& Norway 			& Denmark& Mongolia 			& \\
Finland		& Iceland 			& Malawi & Vietnam  			&	\\
Israel 		& New Zealand 	& Australia & Laos         			&	\\
Japan & & & Afghanistan 		&	 \\
\end{tabular}
\end{table*}

\begin{table*}[htp]
\caption{Countries in the large robust group of the Session 58 voting agreement network (which we highlight in Fig.~\ref{plots113658} using dashed lines).}\label{58table}
\begin{tabular}{l l l}
\multicolumn{3}{c}{North Community}\\
\hline
 United States of America   	& Canada      		& United Kingdom \\
 Ireland 					& Netherlands 		& Belgium \\
 Luxembourg 	 			& France			& Monaco \\
 Liechtenstein 				& Switzerland 		& Spain \\
 Andorra 					& Portugal      		& German Federal Republic \\
 Poland 					& Austria 			& Hungary \\
 Czech Republic 			& Slovakia 		& Italy \\
 San Marino 				& Malta 			& Albania \\
 Macedonia				& Croatia 			& Yugoslavia \\
 Bosnia-Herzegovina 		& Slovenia 		& Greece \\
 Cyprus 					& Bulgaria 		& Moldova \\
 Romania 					& Estonia 			& Latvia \\
 Lithuania 				& Georgia	 		& Finland \\
 Sweden 					& Norway 			& Denmark \\
 Iceland 					& Turkey 			& Israel \\
 South Korea 				& Japan 			& Australia \\
 New Zealand 				& Marshall Islands 	& Palau \\
 Federated States of Micronesia \\
\end{tabular}
\end{table*}

\begin{table*}[htp]
\caption{Countries in the robust groups of the Session 11 network of voting agreements and disagreements (see Fig.~\ref{fig:signed}).
}\label{11signedTable}
\begin{tabular}{l | l | l}
1 & 2 & 3 \\
\hline
Yugoslavia	&United Kingdom	&Poland			\\
Morocco		&Netherlands		&Hungary			\\
Libya		&Belgium			&Czechoslovakia	\\
Sudan		&Luxembourg		&Albania			\\
Egypt		&France			&Bulgaria			\\
Syria			&Portugal			&Romania		\\
Jordan		&South Africa		&Russia			\\
Saudi Arabia	&Israel			&Ukraine			\\
Yemen		&Australia			&Belarus			\\
Afghanistan	&New Zealand		&				\\
India			&				&				\\
Indonesia		&				&				\\
\end{tabular}
\end{table*}

\begin{table*}[htp]
\caption{Countries in the robust groups of the Session 36 network of voting agreements and disagreements (see Fig.~\ref{fig:signed}).}\label{36signedTable}
\begin{tabular}{l l | l}
\multicolumn{2}{l}{1}\vline  & 2 \\
\hline
United States of America \,\,&	German Federal Republic \,\,	&	German Democratic Republic	\\
Canada				&	Italy						&	Poland	\\
United Kingdom		&	Norway					&	Hungary	\\
Ireland				&	Denmark					&	Czechoslovakia	\\
Netherlands			&	Iceland					&	Bulgaria	\\
Belgium				&	Israel					&	Russia	\\
Luxembourg			&	Japan					&	Ukraine	\\
France				&	Australia					&	Belarus	\\
Portugal				&	New Zealand				& 	Mongolia	\\
\end{tabular}
\end{table*}

\begin{table*}[htp]
\caption{Countries appearing in the single robust group of the Session 58 network of voting agreements and disagreements (see Fig.~\ref{fig:signed}).
}\label{58signedTable}
\begin{tabular}{l l l l l}
\hline
Canada			&Spain			&San Marino			&Bulgaria		&Denmark	\\
United Kingdom	&Andorra			&Malta				&Moldova		&Iceland	\\
Ireland			&Portugal			&Albania				&Romania	&Turkey	\\
Netherlands		&Germany		&Macedonia			&Estonia		&South Korea	\\
Belgium			&Poland			&Croatia				&Latvia		&Japan	\\
Luxembourg		&Austria			&Yugoslavia			&Lithuania	&Australia	\\
France			&Hungary			&Bosnia-Herzegovina	&Georgia		&New Zealand	\\
Monaco			&Czech Republic	&Slovenia				&Finland	&		\\
Liechtenstein		&Slovakia			&Greece				&Sweden	&		\\
Switzerland		&Italy			&Cyprus				&Norway	&		\\
\end{tabular}
\end{table*}

\begin{table*}[htp]
\caption{Countries in the three largest robust groups of the Session 11 bipartite network (see Fig.~\ref{fig:bipart}).  We grouped resolutions (not listed) with these countries via community detection on the signed bipartite network of countries and resolutions.}\label{11bipartTable}
\begin{tabular}{l | l | l l l}\label{11bipartiteTable}
1 & 2 & \multicolumn{3}{l}{3} \\
\hline
Poland	&	Finland	&	United States of America	&	Bolivia	&	Sweden	\\
Hungary	&	Ethiopia	&	Canada	&	Paraguay	&	Norway	\\
Czechoslovakia	&	Morocco	&	Cuba	&	Chile	&	Denmark	\\
Albania	&	Tunisia	&	Haiti	&	Argentina	&	Iceland	\\
Bulgaria	&	Libya	&	Dominican Republic	&	Uruguay	&	Liberia	\\
Romania	&	Lebanon	&	Mexico	&	United Kingdom	&	South Africa	\\
Russia	&	Jordan	&	Guatemala	&	Ireland	&	Iran	\\
Ukraine	&	Saudi Arabia	&	Honduras	&	Netherlands	&	Turkey	\\
Belarus	&	Afghanistan	&	El Salvador	&	Belgium	&	Israel	\\
	&	India	&	Nicaragua	&	Luxembourg	&	Taiwan	\\
	&	Myanmar	&	Costa Rica	&	France	&	Pakistan	\\
	&	Sri Lanka	&	Panama	&	Spain	&	Thailand	\\
	&	Nepal	&	Colombia	&	Portugal	&	Laos	\\
	&	Cambodia	&	Venezuela	&	Austria	&	Philippines	\\
	&	Indonesia	&	Ecuador	&	Italy	&	Australia	\\
	&		&	Peru	&	Greece	&	New Zealand	\\
	&		&	Brazil	&		&		\\
\end{tabular}
\end{table*}

\begin{table*}[htp]
\caption{Countries in the four largest robust groups of the Session 36 bipartite network (see Fig.~\ref{fig:bipart}; {resolutions not listed})}\label{36bipartTable}
\scriptsize
\begin{tabular}{l    | l   | l  |lll}
1 & 2 & 3 &\multicolumn{3}{l}{4} \\
\hline
Guatemala	&	Canada	&	Bahamas	&	Cuba	&	Guinea-Bissau	&	Algeria	\\
Spain	&	Ireland	&	Dominican Republic	&	Haiti	&	Gambia	&	Libya	\\
Austria	&	Netherlands	&	Jamaica	&	Trinidad and Tobago	&	Mali	&	Sudan	\\
Greece	&	Belgium	&	Belize	&	Barbados	&	Senegal	&	Iran	\\
Finland	&	Luxembourg	&	Honduras	&	Grenada	&	Benin	&	Iraq	\\
Sweden	&	Portugal	&	Costa Rica	&	St. Lucia	&	Mauritania	&	Syria	\\
Malawi	&	Italy	&	Colombia	&	St. Vincent and the Grenadines	&	Guinea	&	Lebanon	\\
	&	Norway	&	Bolivia	&	Antiqua and Barbuda	&	Burkina Faso	&	Jordan	\\
	&	Denmark	&	Paraguay	&	Mexico	&	Sierra Leone	&	Saudi Arabia	\\
	&	Iceland	&	Chile	&	Nicaragua	&	Ghana	&	Arab Republic of Yemen	\\
	&	Japan	&	Uruguay	&	Panama	&	Togo	&	Peoples Republic of Yemen	\\
	&	Australia	&	Equatorial Guinea	&	Venezuela	&	Cameroon	&	Kuwait	\\
	&	New Zealand	&	Cote D'Ivoire	&	Guyana	&	Nigeria	&	Bahrain	\\
	&		&	Liberia	&	Surinam	&	Chad	&	Qatar	\\
	&		&	Congo	&	Ecuador	&	Congo	&	United Arab Emirates	\\
	&		&	Swaziland	&	Peru	&	Uganda	&	Oman	\\
	&		&	Morocco	&	Brazil	&	Kenya	&	Afghanistan	\\
	&		&	Turkey	&	German Democratic Republic	&	Tanzania	&	Mongolia	\\
	&		&	Myanmar	&	Poland	&	Burundi	&	India	\\
	&		&	Nepal	&	Hungary	&	Rwanda	&	Bhutan	\\
	&		&	Cambodia	&	Czechoslovakia	&	Somalia	&	Pakistan	\\
	&		&	Singapore	&	Malta	&	Djibouti	&	Bangladesh	\\
	&		&	Papua New Guinea	&	Albania	&	Ethiopia	&	Sri Lanka	\\
	&		&	Solomon Islands	&	Yugoslavia	&	Angola	&	Maldives	\\
	&		&	Fiji	&	Cyprus	&	Mozambique	&	Thailand	\\
	&		&	Western Samoa	&	Bulgaria	&	Zambia	&	Laos	\\
	&		&		&	Romania	&	Zimbabwe	&	Vietnam	\\
	&		&		&	Russia	&	Botswana	&	Malaysia	\\
	&		&		&	Ukraine	&	Madagascar	&	Philippines	\\
	&		&		&	Belarus	&	Comoros	&	Indonesia	\\
	&		&		&	Cape Verde	&	Mauritius	&		\\
	&		&		&	Sao Tome \& Principe	&	Seychelles	&		\\
\end{tabular}
\end{table*}


\begin{table*}[htp]
\caption{Countries in the two large robust groups of the Session 58 bipartite network (see Fig.~\ref{fig:bipart}; {resolutions not listed})}\label{58bipartTable}
\scriptsize
\begin{tabular}{l l |  l l l l}
\multicolumn{2}{l}{1} \vline & \multicolumn{4}{l}{2} \\
\hline
Canada	&	Greece	&	Bahamas	&	Senegal	&	Lesotho	&	Myanmar	\\
St. Vincent \& the Grenadines	&	Cyprus	&	Cuba	&	Benin	&	Botswana	&	Sri Lanka	\\
St. Kitts-Nevis	&	Bulgaria	&	Haiti	&	Mauritania	&	Swaziland	&	Maldives	\\
Guatemala	&	Moldova	&	Dominican Republic	&	Niger	&	Madagascar	&	Nepal	\\
Argentina	&	Romania	&	Jamaica	&	Cote D'Ivoire	&	Comoros	&	Thailand	\\
United Kingdom	&	Russia	&	Trinidad and Tobago	&	Guinea	&	Mauritius	&	Cambodia	\\
Ireland	&	Estonia	&	Barbados	&	Burkina Faso	&	Morocco	&	Laos	\\
Netherlands	&	Latvia	&	Dominica	&	Sierra Leone	&	Algeria	&	Vietnam	\\
Belgium	&	Lithuania	&	Grenada	&	Ghana	&	Tunisia	&	Malaysia	\\
Luxembourg	&	Ukraine	&	St. Lucia	&	Togo	&	Libya	&	Singapore	\\
France	&	Armenia	&	Antiqua and Barbuda	&	Cameroon	&	Sudan	&	Brunei	\\
Monaco	&	Georgia	&	Mexico	&	Nigeria	&	Iran	&	Philippines	\\
Liechtenstein	&	Finland	&	Belize	&	Gabon	&	Egypt	&	Indonesia	\\
Switzerland	&	Sweden	&	Honduras	&	Central African Rep.	&	Syria	&	Papua New Guinea	\\
Spain	&	Norway	&	El Salvador	&	Congo	&	Lebanon	&	Vanuatu	\\
Andorra	&	Denmark	&	Nicaragua	&	Dem. Rep.  of Congo	&	Jordan	&	Fiji	\\
Portugal	&	Iceland	&	Costa Rica	&	Uganda	&	Saudi Arabia	&	Nauru	\\
German Fed. Rep.	&	Sao Tome \& Principe	&	Panama	&	Kenya	&	Arab Rep. of Yemen	&	Tonga	\\
Poland	&	Equatorial Guinea	&	Colombia	&	Tanzania	&	Kuwait	&		\\
Austria	&	Chad	&	Venezuela	&	Burundi	&	Bahrain	&		\\
Hungary	&	Turkey	&	Guyana	&	Rwanda	&	Qatar	&		\\
Czech Republic	&	Tajikistan	&	Surinam	&	Somalia	&	United Arab Emirates	&		\\
Slovakia	&	Uzbekistan	&	Ecuador	&	Djibouti	&	Oman	&		\\
Italy	&	Kazakhstan	&	Brazil	&	Ethiopia	&	Afghanistan	&		\\
San Marino	&	South Korea	&	Bolivia	&	Eritrea	&	Turkmenistan	&		\\
Malta	&	Japan	&	Paraguay	&	Angola	&	China	&		\\
Albania	&	Australia	&	Belarus	&	Mozambique	&	Mongolia	&		\\
Macedonia	&	New Zealand	&	Azerbaijan	&	Zambia	&	North Korea	&		\\
Croatia	&	Solomon Islands	&	Cape Verde	&	Zimbabwe	&	India	&		\\
Yugoslavia	&	Kiribati	&	Guinea-Bissau	&	Malawi	&	Bhutan	&		\\
Bosnia-Herzegovina	&	Tuvalu	&	Gambia	&	South Africa	&	Pakistan	&		\\
Slovenia	&	Western Samoa	&	Mali	&	Namibia	&	Bangladesh	&		\\
\end{tabular}
\end{table*}

\clearpage{}

\end{document}